\documentclass[%
 reprint,
 amsmath,amssymb,
 aps,
]{revtex4-1}
\usepackage[colorlinks=true,linkcolor=blue]{hyperref}
\usepackage{graphicx}
\usepackage{dcolumn}
\usepackage{bm}

\begin{document}

\title{Effects of the variation of
mass on fermion localization and resonances on thick branes}

\author{Zhen-Hua Zhao}
 \email{zhaozhh09@lzu.cn}
\author{Yu-Xiao Liu}%
 \email{Corresponding author. liuyx@lzu.edu.cn}
\author{Hai-Tao Li} \email{liht07@lzu.cn} \author{Yong-Qiang Wang} \email{yqwang@lzu.edu.cn}
\affiliation{Institute of Theoretical Physics,
    Lanzhou University, Lanzhou 730000,
    People's Republic of China \\}

\begin{abstract} A few years ago, Campos investigated the critical phenomena of thick branes in
warped spacetimes [Phys. Rev. Lett. {\bf 88} (2002) 141602]. Inspired by his work, we consider a toy
model of thick branes generated by a real scalar field with the potential
$V(\phi)=a\phi^2-b\phi^4+c\phi^6$, and investigate the variation of the mass parameter $a$ on the branes as well as
the localization and resonances of fermions. An interesting result is found:
{there is a critical value for the mass parameter $a$}, and when the critical
  {value of $a$} is reached the solution of the background scalar field is not unique and has the shape
of a double kink. This happens in both cases with and without gravity. It is also shown that the
numbers of the bound Kaluza-Klein modes of fermions on the gravity-free brane and the resonant states
of fermions on the brane with gravity increase with the   {value of $a$}. \end{abstract}


 \maketitle

\section{Introduction}
The idea of embedding our Universe in a higher dimensional space-time provides new insights for solving some
long-standing problems in physics, such as the problems of the gauge hierarchy \cite{Antoniadis1998436,Arkani-Hamed1998429,Randall199983} and
cosmological constant~\cite{Rubakov1983}. Since the discovery of four-dimensional gravity
can be realized on a brane in five-dimensional (5D) space-time~\cite{Randall199983a},
the brane-world theories have received considerable attention.
The branes in Randall and Sundrum model \cite{Randall199983,Randall199983a} are fixed on some points along the fifth dimension,
 with a form of $\delta$ function. The Randall and Sundrum brane-world model is very ideal,
 it is infinitely thin, and the formation of it lacks the dynamical mechanism.
 A natural extension of thin brane models is the thick brane scenarios with the gravity coupled to
five-dimensional scalar fields \cite{Bonjour1999,Goldberger1999,Csaki2000581,DeWolfe200062,Gremm2000478,Ichinose200118,
Ichinose200118a,Kehagias2001504,Campos200288,Gregory2002,Gregory2002a,Kobayashi2002,
Bazeia200391,Bronnikov2003,Eto200368,Melfo2003,Bazeia2004,Bazeia200405,Bazeia2006f,
Dzhunushaliev2006,Dzhunushaliev200713,George2007,Dzhunushaliev200877,
Bazeia2009c,Burnier2009,Chumbes2010,Dzhunushaliev200979,Liu2009c}.
Furthermore, there are some other thick brane models such as the models with thick branes arising from
pure geometry \cite{Arias2002643,Barbosa-Cendejas200510,Barbosa-Cendejas200673,Barbosa-Cendejas2007, Barbosa-Cendejas200877,Liu2009g}
or fermion self-interaction \cite{Andrianov2003, Andrianov2005}.
Some reviews about thick brane-world models can be found in Refs.~\cite{SenGupta2008,Dzhunushaliev2009,Shifman2009,Rizzo2010}.
In brane-world theories, possessing the possibility of localizing all kinds of matter fields on branes is very important.
The localization of particles  on a thick brane without gravity was suggested a long time ago \cite{Rubakov1983}.
In the case with gravity, the localization of fermions and bosons on various brane worlds has been studied in much of the literature
\cite{Bajc2000,Randjbar-Daemi2000,Pomarol2000,Oda2001,Akhmedov2001,
Dvali2001,Ghoroku2002,Ringeval200265,
Koley200522,Wang2005,Melfo2006,Liu2007a,Liu2007,Davies2007,Slatyer2007,Koley2008,
Zhang2008,Liu200878,Liu200808,Liu200802,Guerrero2009,Zhao2009,
Liu2009a,Liu2010}. Furthermore, for some thick branes, the massive fermions trapped on them are unstable and have a finite confinement lifetime~\cite{Ringeval200265,Davies2007,Dubovsky200062,Almeida2009,Liu2009,Liu2009a,Liu2009f,Liu2010a}.

Recently, Campos discussed the critical phenomena of thick branes in warped spacetimes~\cite{Campos200288}. In his work, the thick branes are constructed with a complex scalar field, and the real part of the complex scalar field is used as an order parameter field to describe the first-order phase transition.   {In his paper the potential of the background complex scalar field $\Phi$ has the form of
\begin{equation}
V(\Phi)=A|\Phi|^2-B\phi_R(\phi^2_R-3\phi_I^2)+C|\Phi|^4,
\end{equation}
where $\phi_R$ and $\phi_I$ are the real and the imaginary parts of the complex scalar field $\Phi$, respectively. Taking the effect of the temperature into account, the mass parameter $A$  should be a function of the temperature $T$ \cite{Dolan1974,Weinberg1974,Jackiw1975}}. Campos  found that the presence of gravity would lower the critical temperature of the phase transition corresponding to what occurred in Euclidean space~\cite{Campos200288}. And when the critical temperature is reached an interface that interpolates two bulk ordered phases splits into two interfaces with a disordered phase between them~\cite{Campos200288}.

Inspired by the work of Campos, in this paper, we consider the thick branes with a real scalar field as the order parameter and the background field. For the purpose of investigating the universal property of the model, the potential of the scalar field is taken as
\begin{eqnarray}
  V(\phi)=a\phi^2-b\phi^4+c\phi^6,\label{Vphi6}
\end{eqnarray}
where $a, b, c>0$.
Taking the effect of the temperature into account, different from the $\phi^4$ model, then $a$ and $b$ should all be functions of temperature $T$ {\cite{Joseph1982,Su198720}}. But to calculate the exact forms of $a(T)$ and $b(T)$ is very difficult in the five-dimensional curved space-time, although a five-dimensional field in the bulk can be regarded as a four-dimensional field on the brane with an infinite Kaluza-Klein (KK) tower of masses   { \cite{Brevik2001599,Bazeia200411}}. In brane-world theories, we have not found the exact form of the effective potential of the $\phi^6$ model with temperature corrections.

Without the exact forms of $a(T)$ and $b(T)$, here we just investigate the effects of   {the variation of the mass parameter $a$} on the fermion localization and resonances.  {This can partly represent the effects of the temperature on the branes in this $\phi^6$ model}. In our investigation, we find that the number of the bound states or {of} the resonant states increases with the   {parameter $a$}. Besides, when the critical   {value of $a$} is reached the solution of the background scalar field is not unique and it has the shape of a double kink, which happens in the two cases with and without gravity. That means when the system approaches the critical   {value of $a$} the width of the brane will become uncertain.


The paper is organized as follows. In Sec.~\ref{sec:gravity-free},   {the effects of the variation of the mass parameter $a$} on thick branes are discussed in the case without gravity. This includes the effects on the solution of the background scalar field and on fermion localization. In order to localize fermions on the branes, the Yukawa coupling between the background scalar field and fermions is introduced. Then Sec.~\ref{sec:wgravity} deals with the case of thick branes with gravity. For the same Yukawa coupling, the gravity on the branes will lead to the disappearance of the discrete spectrum of fermions, and only one bound state of fermions, i.e., the four-dimensional massless fermion, resides on the branes. Furthermore, fermion resonances with a finite lifetime could appear on the branes, and their number increases with the   {mass parameter $a$}. Finally, a brief conclusion is presented in Sec.~\ref{sec:conclusion}.

\section{Without gravity}\label{sec:gravity-free}

\begin{figure}
\begin{center}
\includegraphics[width=0.4\textwidth]{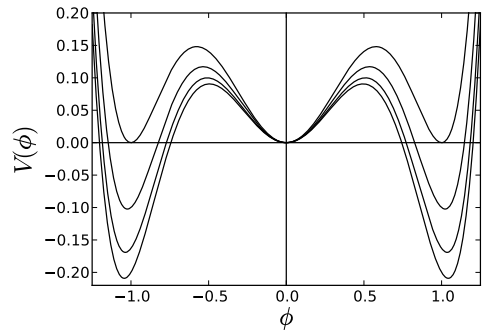}
\end{center}
\caption{ \label{fig:Vphi}
The profiles of the potential as a function of $\phi$ with $b=2$ and $c=1$ ($a_c=1$). The lines correspond to $a=0.8, 0.837, 0.9, 1$ from bottom to top, respectively.}
\end{figure}

We begin with the case without gravity. Consider a 5D model containing one real scalar field. The action is
\begin{eqnarray}
       S = \int d^4x dy
           \left[ - \frac{1}{2} \eta^{MN}\partial_M \phi \partial_N \phi
             - V(\phi)\right],\label{action1}
\end{eqnarray}
where $M, N=(0,1,2,3,5)$ and $\eta^{MN}=\text{diag}(-1, 1, 1, 1, 1)$. The self-interaction potential of the background scalar field is of the form as (\ref{Vphi6}). There are three local minima for $V(\phi)$, one is at $\phi^{(1)}=0$ corresponding to a disordered bulk phase and the other two are at $\phi^{(2)}=-\phi^{(3)}=v$ with
 \begin{equation}
 v=\sqrt{\frac{\sqrt{b^2-3 a c}}{3 c}+\frac{b}{3 c}} \;, \label{v}
 \end{equation}
they are degenerated and correspond to ordered bulk phases. When
$a= a_c$~[$a_c=b^2/4c$], $V(\phi^{(1)})=V(\phi^{(2)})=V(\phi^{(3)})$,
$V(\phi)$ has three degenerated global minima.
So we take $a_c$ as the parameter corresponding
to the critical value of   {$a$}~\cite{Campos200288}.
The profiles of the potential as a function
of the $\phi$ are shown in Fig.~{\ref{fig:Vphi}}.
This is very similar to the model in Ref.~\cite{Campos200288}, where the thick branes are constructed from a complex scalar field and the action has $\mathbb{Z}_3$ symmetry. The usual hypothesis is that $\phi$ is only the function of the extra dimension $y$, i.e., $\phi=\phi(y)$.

\subsection{Effects of the variation of the mass parameter $a$ on the thick brane solutions} \label{subsec:Branegf}

The equation of motion for $\phi$ corresponding to the action (\ref{action1}) reads as
\begin{eqnarray}
\phi''=\frac{d V(\phi)}{d \phi},\label{Eq:Phi}
\end{eqnarray}
where the prime stands for the derivative with respect to $y$. When $y\rightarrow\pm\infty$, $\phi(\pm\infty)={\pm} v$ and $\phi'(\pm\infty)=0$. Considering the symmetry of the potential $V(\phi)$, we can set $\phi(0)=0$.
In summary, the boundary conditions of Eq. (\ref{Eq:Phi}) are
\begin{eqnarray}
\phi(+\infty)=-\phi(-\infty)=v ~\text{and}~\phi(0)=0.\label{eq:boundary1}
\end{eqnarray}

\begin{figure}
\includegraphics[width=0.4\textwidth]{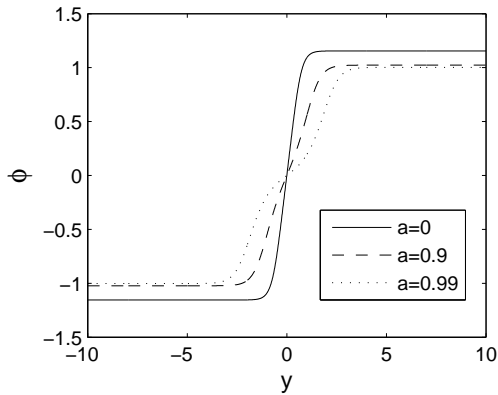}
\caption{ \label{fig:Phi_gravity-free}
The profiles of the background scalar field $\phi$ in the gravity-free case with $b=2$ and $c=1$ ($a_c=1$). The three lines correspond to $a=0$, $a=0.9$, and $a=0.99$, respectively.}
\end{figure}

\begin{figure}
\includegraphics[width=0.4\textwidth]{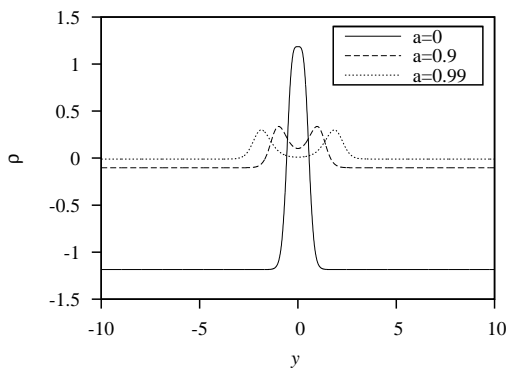}
\caption{ \label{fig:energyDensityOfPhiWithoutGravity}
The profiles of the energy density for the background scalar field in the gravity-free case with $b=2$ and $c=1$ ($a_c=1$). The three lines correspond to $a=0$, $a=0.9$, and $a=0.99$, respectively.}
\end{figure}

We have numerically solved the equation (\ref{Eq:Phi}) using the above boundary conditions with $b=2$, $c=1$ and different  {values of $a$ with} $a<a_c=1$; the results are shown in Fig.~{\ref{fig:Phi_gravity-free}}. From the figure we can find that with the increase of  {the value of $a$} the profile of the background field $\phi$ has the tendency to hold the form of a double kink. For the energy density of the background scalar field, we have
\begin{eqnarray}
\rho=\frac{1}{2}\phi'^2+V(\phi).
\end{eqnarray}
Its profiles with different values of $a$ are shown in Fig.~{\ref{fig:energyDensityOfPhiWithoutGravity}}. We see that for $a=0$, the energy density has a single-peak around $y=0$. As  the value of $a$ is increasing, the maximum of $\rho$ splits into two new maxima, and the distance of the new maxima increases with  {the value of a}. That means the brane gets thicker and thicker with the increase of  {the value of a}.

\begin{figure}
\includegraphics[width=0.4\textwidth]{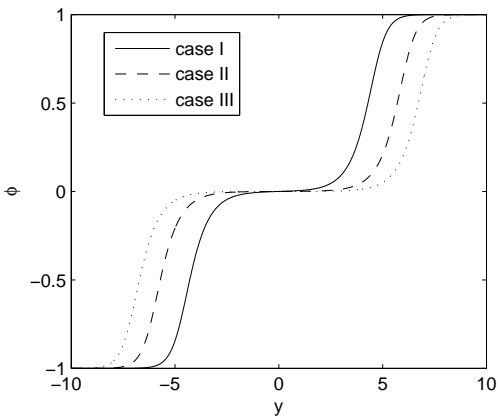}
\caption{ \label{fig:Phi_gravity-free_at_ac}
The profiles of the background scalar field $\phi$ in the gravity-free case with $a=a_c$. The three lines correspond to three numerical solutions with different initial guess solutions.}
\end{figure}

When $a=a_c$ we find that the profile of $\phi$ has a double-kink form and the solution of $\phi$ is not unique\footnote{
Because Eq. (\ref{Eq:Phi}) is non-linear, so in order to solve it numerically we need to offer an initial guess solution of $\phi$.  And different guess solutions follow different solutions of $\phi$.}, which means that the boundary conditions (\ref{eq:boundary1}) can not determine the solution uniquely. Figure~\ref{fig:Phi_gravity-free_at_ac} shows the profiles of the different solutions of $\phi$ with $a=a_c$. This is very different from the noncritical case. So in the brane-world model, when the critical  {value of $a$} is reached the ordered-ordered brane (located between two ordered phases) splits into two ordered-disordered interfaces (with each one located between an ordered phase and a disordered phase). The phase transition can be identified by $\phi'(0)=0$.

\subsection{Fermions on the brane without gravity} \label{subsec:fwithoutg}

In order to localize fermions on the brane, we need to introduce the Yukawa coupling of the fermion with the background scalar field $\phi$.
The action of the bulk fermion is
\begin{eqnarray}
 S_{\Psi}=\int d^5x \Big[\bar\Psi\gamma^M\partial_M\Psi-\eta\bar\Psi
 \phi\Psi\Big],
\end{eqnarray}
where $\eta$ is the coupling constant, $\gamma^M=(\gamma^\mu, \gamma^5)$ $(\mu=0,1, 2, 3)$ with $\gamma^\mu$ the usual 4D Dirac matrices and $\gamma^5$ the chirality operator. From the above action we can obtain the equation of motion for the fermion
\begin{eqnarray}
 \Big[\gamma^{\mu}\partial_{\mu}+\gamma^5\partial_y-\eta
\phi\Big]\Psi=0.
\end{eqnarray}
In order to solve this equation, we need to separate variables with the usual KK and chiral decomposition:
\begin{eqnarray}
 \Psi(x,y)=\sum_n \Big(\psi_{{\rm{L}} n}(x)f_{{\rm{L}} n}(y)
 +\psi_{{\rm{R}} n}(x)f_{{\rm{R}} n}(y)\Big),
\end{eqnarray}
where the sum also includes the integral over all continuum cases. The 4D left- and right-handed fermions $\psi_{{\rm{L}} n}$ and $\psi_{{\rm{R}} n}$ satisfy the following coupled equations:
\begin{eqnarray}
\gamma^{\mu}\partial_{\mu}\psi_{{\rm{L}} n}&=&m_{n}\psi_{{\rm{R}} n},\\
\gamma^{\mu}\partial_{\mu}\psi_{{\rm{R}} n}&=&m_{n}\psi_{{\rm{L}} n}.
\end{eqnarray}
The KK modes $f_{{\rm{L}} n}$ and $f_{{\rm{R}} n}$ satisfy
\begin{eqnarray}
 f_{{\rm{L}} n}'+\eta\phi f_{{\rm{L}} n}&=& m_n f_{{\rm{R}} n},\label{eq:NoGFL}\\
 f_{{\rm{R}} n}'-\eta\phi f_{{\rm{R}} n}&=&-m_n f_{{\rm{L}} n},\label{eq:NoGFR}
\end{eqnarray}
and the following orthonormality conditions:
\begin{eqnarray}
 \int_{-\infty}^{\infty} f_{{\rm{L}} m} f_{{\rm{L}} n}dz
   &=&\delta_{mn}=\int_{-\infty}^{\infty} f_{{\rm{R}} m} f_{{\rm{R}} n}dz,\nonumber\\
    \int_{-\infty}^{\infty} f_{{\rm{L}} m} f_{{\rm{R}} n}dz&=&0.
\end{eqnarray}
From Eqs. (\ref{eq:NoGFL}) and (\ref{eq:NoGFR}) we can get the following  Schr\"{o}dinger equations with eigenvalue $m_n^2$:
\begin{subequations}\label{eq:Sch}
\begin{eqnarray}
  &&\big[-\partial_y^2+V_{\rm{L}}(y)\big]f_{{\rm{L}} n}
    = m^2_n f_{{\rm{L}} n},\\
  &&\big[-\partial_y^2+V_{\rm{R}}(y)\big]f_{{\rm{R}} n}
    = m^2_n f_{{\rm{R}} n},
\end{eqnarray}
\end{subequations}
where the potentials are
\begin{eqnarray}
  &&V_{\rm{L}}(y)=\eta^2\phi^2-\eta\phi' ,\\
  &&V_{\rm{R}}(y)=\eta^2\phi^2+\eta\phi'.
\end{eqnarray}

Before solving the Schr\"{o}dinger Eqs.~(\ref{eq:Sch}), let us take a look at the behavior of $V_{\rm{L}}(y)$ and $V_{\rm{R}}(y)$ at $y=0$ and $y\rightarrow \pm\infty$.

We first focus on the case of $a<a_c$. With the boundary condition $\phi(0)=0$, we have $V_{\rm{L}}(0)=-\eta\phi'(0)$ and $V_{\rm{R}}(0)=+\eta\phi'(0)$. From the numeric solutions in Fig.~{\ref{fig:Phi_gravity-free}}, we can find that $\phi'(0)>0$. So in the case of $\eta>0$, $V_{\rm{L}}(0)<0$ and $V_{\rm{R}}(0)>0$. Without loss of generality we choose $\eta>0$.
When $y\rightarrow \pm\infty$, $\phi(\pm\infty)=\pm v$ and $ {\phi'(\pm\infty)}=0$, so $V_{{\rm{L}},{\rm{R}}}(\pm\infty)=\eta^2 v^2$, which is a constant greater than zero.

\begin{figure*}
\includegraphics[width=0.4\textwidth]{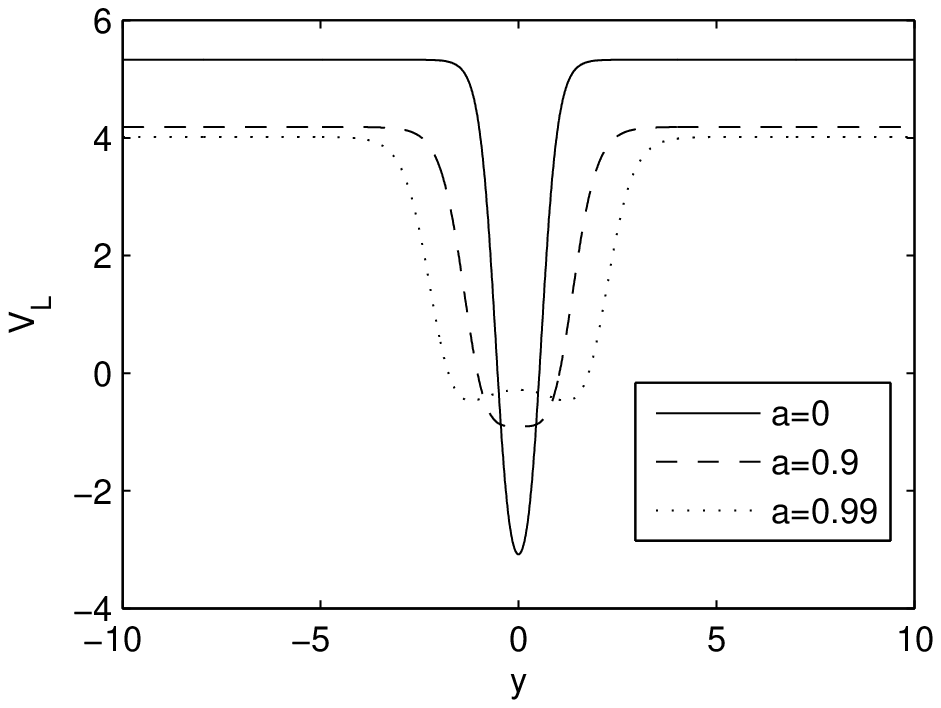}
\includegraphics[width=0.4\textwidth]{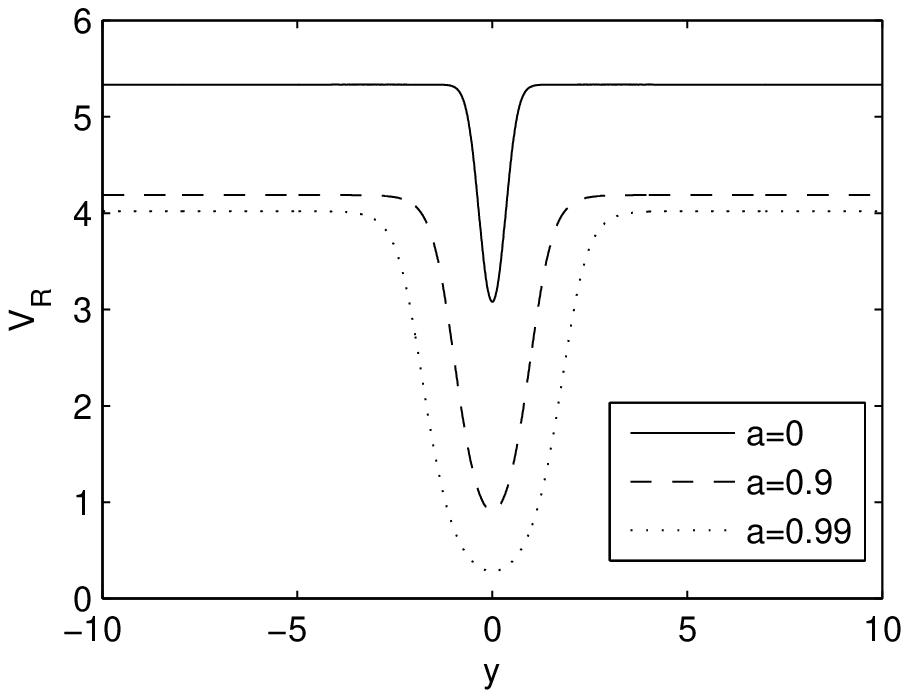}
\caption{ \label{fig:Potential_grvity-free}
The profiles of the potentials $V_{\rm{L}}$ and $V_{\rm{R}}$. The parameters are set to $b=2,c=1,\eta=2$, and $a=0,0.9,0.99$. }
\end{figure*}

The shapes of the potentials $V_{\rm{L}}$ and $V_{\rm{R}}$ for different values of  {$a$} are shown in Fig.~\ref{fig:Potential_grvity-free}. From the figure, we can find: (a) The potential  {of} the left-handed fermion has a negative potential well, so the left fermion may has a massless mode. (b) As $a$ is close to the critical  {value} there will be two minima in the potential well. (c) For the right-handed fermion case, the potential $V_{\rm{R}}$ has also a finite well when $\eta>\phi'(0)/v$ and does not have any potential well when $0<\eta\leq\phi'(0)/v$, but the minimum of the potential is always greater than zero for any $\eta>0$, so there will not exist a zero mode.

Here, it should be noted that the form of the potential well is
also related to other parameters such as $b$, $c$, and $\eta$. But
in this paper we lay the emphasis on the effects of the  {variation of mass parameter $a$} on the branes, so the effects of
other parameters will not be taken into account.

We have solved Eqs. (\ref{eq:Sch}) numerically for the case of the  {value of $a$ below its} critical value. And we get the mass spectra of the left- and right-handed fermions, which are shown in Tables~\ref{tab:mass_grvity-free-left} and~\ref{tab:mass_grvity-free-right}, respectively. From the two tables we find: (a) The number of the bound KK modes of both the left- and right-handed fermions increases with  {the value of $a$}. (b) The spectra of the massive KK modes of the left- and right-handed fermions are almost the same, which demonstrates that a Dirac fermion could be composed from the left- and right-handed bound KK modes \cite{Liu2009}.

For the case of $a=a_c$, because the solution of the background scalar  {field $\phi$}
is not unique and there are no other conditions to fix it,
so the localization of fermions in this case is not taken into account here.

\begin{table}
\begin{tabular}{|c|c|c|c|c|c|}
  \hline
   $a$  &The maximum of $V_{\rm{L}}$&$m_0^2$ & $m_1^2$   & $m_2^2$  & $m_3^2$\\
   \hline
  $0$   &$5.333333$         &$0 $    & $4.797808$&   & \\
  \hline
  $0.5$ &$4.774852$         &$0 $    & $3.748817$&  & \\
  \hline
  $0.9$ &$4.1869012$        &$0 $    & $2.118862$&$4.003805$&\\
  \hline
  $0.99$&$4.019852$         &$0 $    & $0.995953$&$2.496547$&$3.824792$\\
  \hline
\end{tabular}
\caption{ \label{tab:mass_grvity-free-left}
The mass spectrum of the left-handed fermion in the gravity-free case with different values of $a$ and fixed $b=2$, $c=1$, and $\eta=2$.}
\end{table}

\begin{table}
\begin{tabular}{|c|c|c|c|c|c|}
  \hline
  $a$   & The maximum of $V_{\rm{R}}$ & $m_0^2$    & $m_1^2$    & $m_2^2$\\
   \hline
  $0$   &$5.333333$           & $4.797808$ &         & \\
  \hline
  $0.5$ &$4.774852$           & $3.748817$ &         & \\
  \hline
  $0.9$ &$4.186901$           & $2.118856$ & $4.003659$ & \\
  \hline
  $0.99$&$4.019852$           & $0.994289$ & $2.496545$ & $3.824551$\\
  \hline
\end{tabular}
\caption{ \label{tab:mass_grvity-free-right}
The mass spectrum of the right-handed fermion in the gravity-free case with different values of $a$ and fixed $b=2$, $c=1$, and $\eta=2$.}
\end{table}

\section{With gravity} \label{sec:wgravity}

In the scenario of the thick brane world with gravity, the action is as follows:
\begin{equation}
       S = \int d^4x dy \sqrt{-g}
           \left(  \frac{1}{4} R - \Lambda
             - \frac{1}{2} g^{MN} \partial_M \phi
                                      \partial_N \phi
             - V(\phi)
           \right),\label{action2}
\end{equation}
where $\Lambda$ is the 5D bulk cosmological constant, and the scalar potential $V(\phi)$ is the same as (\ref{Vphi6}).
The metric in this model is assumed as
\begin{eqnarray}
  ds^2=e^{2A(y)}\eta_{\mu\nu}dx^\mu dx^\nu+dy^2,\label{metric}
\end{eqnarray}
where $\mu, \nu=0,1,2,3$, $\eta_{\mu\nu}=\text{diag}(-1,1,1,1)$, and $e^{2A(y)}$ is the warp factor. {The} usual hypothesis is that $A$  is only the function of the extra dimension $y$.

\subsection{ {Effects of the variation of the mass parameter $a$ on} the thick brane solutions}
\label{subsec:brane}
The equations generated from the action (\ref{action2}) with ansatz (\ref{metric}) reduce to the following coupled nonlinear differential
equations
\begin{eqnarray}
\phi''&=&-4A' \phi'+\frac{d V(\phi)}{d \phi},\label{EqM1}\\
A''&=&-\frac{2}{3}\phi'^{2},\label{EqM2}\\
A'^{2}&=&\frac{1}{6}(\phi'^{2}-2V)-\frac{\Lambda}{3}.\label{EqM3}
\end{eqnarray}
Equation~(\ref{EqM3}) can be used to fix the bulk cosmological constant $\Lambda$.
When $y\rightarrow\pm\infty$, $|\phi(\pm\infty)|= \phi^{(2)}=-\phi^{(3)}=v$. In this paper we chose $\phi(+\infty)=v$ and $\phi(-\infty)=-v$.
The boundary conditions can be read as follows \cite{Campos200288}:
\begin{subequations}
\label{bdcsz}
\begin{eqnarray}
A(0)=A'(0)=\phi(0)=0,\\
\phi(+\infty)=-\phi(-\infty)=v.
\end{eqnarray}
\end{subequations}

\begin{figure*}
\includegraphics[width=0.4\textwidth]{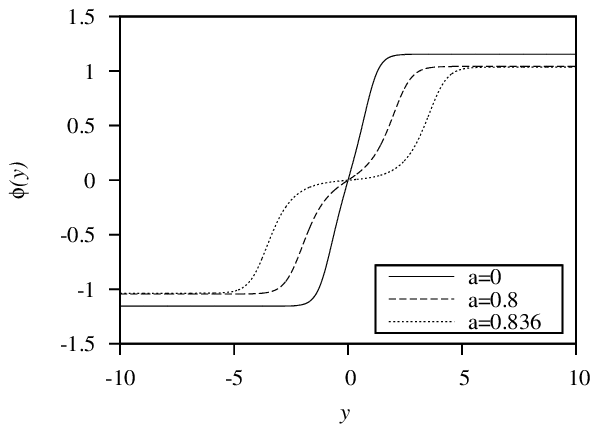}
\includegraphics[width=0.4\textwidth]{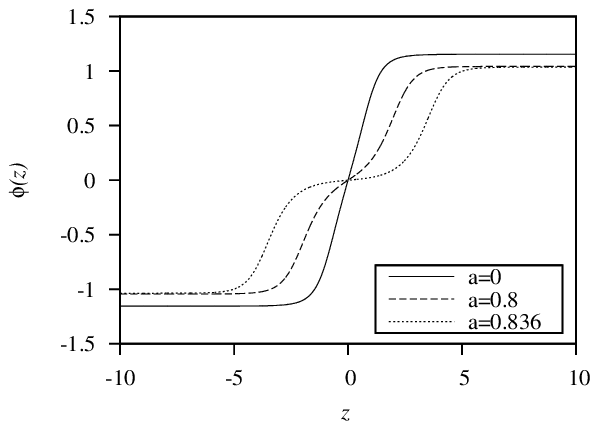}
\caption{ \label{fig:Phi}
The profiles of the background scalar field $\phi$. The parameters are set to $b=2,c=1$, and $a=0,0.8,0.836$.}
\end{figure*}
\begin{figure*}
\includegraphics[width=0.4\textwidth]{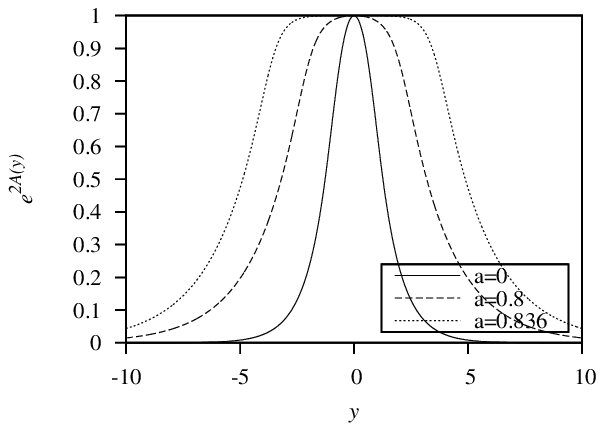}
\includegraphics[width=0.4\textwidth]{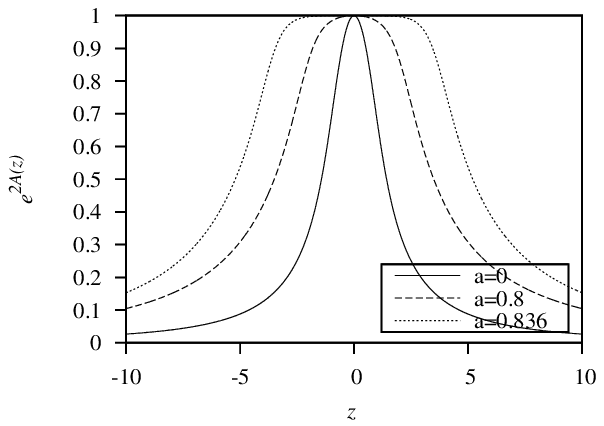}
\caption{ \label{fig:WarpFactor}
The profiles of the warp factor $e^{2A}$. The parameters are set to $b=2,c=1$, and $a=0,0.8,0.836$.}
\end{figure*}

\begin{figure*}
\includegraphics[width=0.4\textwidth]{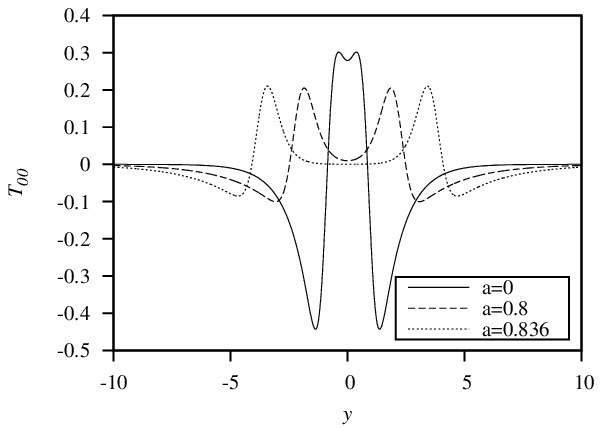}
\includegraphics[width=0.4\textwidth]{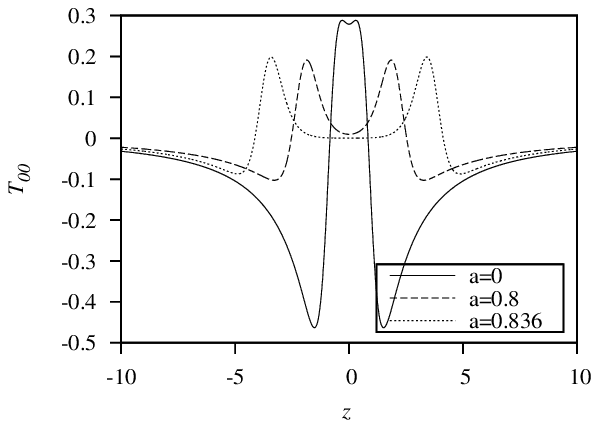}
\caption{ \label{fig:T00}The profiles of the matter energy density $T_{00}$. The parameters are set to $b=2,c=1$, and $a=0,0.8,0.836$.}
\end{figure*}

With the boundary conditions (\ref{bdcsz}) we have solved Eqs. (\ref{EqM1}) and (\ref{EqM2}) numerically.
The results are shown in Figs.~{\ref{fig:Phi}} and \ref{fig:WarpFactor}.
And the profiles of the matter energy density
\begin{eqnarray}
T_{00}=e^{2A}(\frac{1}{2}\phi'^{2}+V(\phi))
\end{eqnarray}
are shown in Fig.~{\ref{fig:T00}}.  {The effect of the variation of the mass parameter $a$ on the} energy density is similar to the case without gravity.
Through the comparison of Figs.~{\ref{fig:energyDensityOfPhiWithoutGravity}} and~{\ref{fig:T00}} we see that the existence of gravity can change the behavior of $T_{00}$ at infinity.  When $y\rightarrow \pm \infty$, $T_{00}$ goes to zero in the case with gravity, but in the case without gravity it goes to a negative constant.

In the case without gravity, we can determine easily the value of the critical  {value of $a$}, which turns out to be $a_c=b^2/(4c)$.
But in the case of existence of gravity, just like the discussion in Ref.~\cite{Campos200288}, the critical  {value of $a$} is not $a_c$ but a smaller effective critical {value} $a_*$. The reason why $a_*<a_c$ can be interpreted as follows: When $a=a_*$ the thick brane will coexist in three phases, in other words, the solution of the background scalar $\phi$  has the form of a double kink, which means $\phi'(0)=0$. Then from Eq.~(\ref{EqM3}) with the conditions (\ref{bdcsz}) we find this leads to the result that the bulk cosmological constant $\Lambda=0$. So when $y\rightarrow\infty$, $A'^{2}(\infty)=-V_{\text{min}}(a_*)/3$ ($V_{\text{min}}(a)=V(\phi(\infty))$ is the global minima of the scalar potential with fixed $a$. Because of the existence of gravity, i.e., $A'^{2}(\infty)>0$, we have $V_{\text{min}}(a_*)<0$. Thus, noting that $V_{\text{min}}(a)$ increases with $a$ and $V_{\text{min}}(a_c)=0$, we reach the conclusion $a_*<a_c$.

We also find just like the case without gravity when $a=a_*$ the solution of $\phi$ is not unique. For $b=2$ and $c=1$, the critical  {value of $a$} is $a_*=0.837$.

\subsection{Fermions on the brane with gravity}\label{subsec:fwithg}
The localization of fermions on the brane with gravity is very similar to the case without gravity. But in order to get the mass-independent potential for KK modes of fermions, we will follow Refs.~\cite{Randall199983,Randall199983a}
to change the metric (\ref{metric}) to a conformally flat one
\begin{eqnarray}
ds^2=e^{2A}(\eta_{\mu\nu}dx^{\mu}dx^{\nu}+dz^2),\label{metric2}
\end{eqnarray}
by performing the coordinate transformation
\begin{eqnarray}
dz=e^{-A(y)}dy.\label{eq:relation}
\end{eqnarray}
The Dirac action of a massless 5D spin $1/2$ fermion coupled to the scalar and gravity is
\begin{eqnarray}
 S_{\Psi}=\int dx^5\sqrt{-g}\Big\{\bar\Psi\Gamma^M(\partial_M+\omega_M)\Psi-\eta\bar\Psi
 \phi\Psi\Big\},
 \end{eqnarray}
where $\Gamma^M=(e^{-A}\gamma^{\mu},e^{-A}\gamma^{\mu})$ are the curved space gamma matrices. $\omega_M$ are the spin connection \cite{Duan1958,Fischbach1981, Zhao200776} and their nonvanishing components are \cite{Liu2009f, Ringeval200265}
\[
\omega_{\mu}=\frac{1}{2}A'\gamma_{\mu}\gamma_{5}.
\]
where the prime denotes the derivative with respect to $z$.
Using the conformal metric (\ref{metric2}), the resulting Dirac
equation is
\begin{eqnarray}
 \Big[\gamma^{\mu}\partial_{\mu}+\gamma^5(\partial_z+2 A')-\eta
e^{A}\phi\Big]\Psi=0.
\end{eqnarray}
With the usual KK and chiral decomposition
\begin{eqnarray}
 \Psi(x,z)=\sum_n \Big(\psi_{{\rm{L}} n}(x)f_{{\rm{L}} n}(z)
 +\psi_{{\rm{R}} n}(x)f_{{\rm{R}} n}(z)\Big),
\end{eqnarray}
we can obtain the coupled equations for $f_{{\rm{L}} n}$ and $f_{{\rm{R}} n}$
\cite{Liu200878}:
\begin{eqnarray}
\big[{\partial_{z}+2 A'+\eta e^A
F(\phi,\chi)}\big]f_{{\rm{L}} n}(z)&=&m_n f_{{\rm{R}} n}(z),\label{eq:zeroMode}\\
\big[{\partial_{z}+2A'-\eta e^A
F(\phi,\chi)}\big]f_{{\rm{R}} n}(z)&=&-m_n f_{{\rm{L}} n}(z),~~~~~
\end{eqnarray}
where $f_{{\rm{L}} n}$ and $f_{{\rm{R}} n}$ satisfy the following orthonormality conditions:
\begin{eqnarray}\label{orthonormalityConditions}
 \int_{-\infty}^{\infty}e^{4A}f_{{\rm{L}} m}f_{{\rm{R}} n}dz
    =\delta_{m n}\delta_{{\rm{L}} {\rm{R}}}.
\end{eqnarray}
Defining $f_{{\rm{L}}}=\tilde f_{{\rm{L}}}e^{-2A}$
and $f_{{\rm{R}}}=\tilde f_{{\rm{R}}}e^{-2A}$, we obtain the
Schr\"{o}dinger-like equations for the new functions
$\tilde f_{{\rm{L}}}$ and $\tilde f_{{\rm{R}}}$ \cite{Liu200878}:
\begin{subequations}
\label{eq:motionF}
\begin{eqnarray}
  &&\big[-\partial_z^2+V_{\rm{L}}(z)\big]\tilde f_{{\rm{L}} n}
    = m^2_n\tilde f_{{\rm{L}} n},\\
  &&\big[-\partial_z^2+V_{\rm{R}}(z)\big]\tilde f_{{\rm{R}} n}
    = m^2_n\tilde f_{{\rm{R}} n},
\end{eqnarray}
\end{subequations}
where the effective potentials are given by
\begin{subequations}\label{eq:V}
\begin{eqnarray}
  V_{{\rm{L}}}(z)&=& \eta^2 e^{2A}\phi^2-\eta e^A \phi'-\eta \phi e^A A' ,\\
  V_{{\rm{R}}}(z)&=& \eta^2 e^{2A}\phi^2+\eta e^A\phi'+\eta \phi e^A A'.
\end{eqnarray}
\end{subequations}
The equations of motion of the background scalar field $\phi$ and $A$ in the $z$ coordinate with metric (\ref{metric2})  can be written as
\begin{subequations}\label{eq:eqmz}
\begin{eqnarray}
\phi''&=&-3A' \phi'+e^{2A}\frac{d V(\phi)}{d \phi},\label{EqMz1}\\
A''&=&A'^{2}-\frac{2}{3}\phi'^{2},\label{EqMz2}\\
A'^{2}&=&\frac{1}{6}\phi'^{2}-\frac{1}{3}e^{2A}(V(\phi)+\Lambda).\label{EqMz3}
\end{eqnarray}
\end{subequations}
Before solving numerically the Schr\"{o}dinger equations~(\ref{eq:motionF}), let us analyze the behavior of $V_{\rm{L}}(z)$ and ${V_{\rm{R}}}(z)$ at $z=0$ and $z\rightarrow \pm\infty$.

At $z=0$, $A(0)=A'(0)=\phi(0)=0$, so $V_{\rm{L}}(0)=-\eta\phi'(0)$ and $V_{\rm{R}}(0)=\eta\phi'(0)$. From the numeric solutions of $\phi(z)$ in Fig.~{\ref{fig:Phi}}, we can find $\phi'(0)>0$. Then for $\eta>0$ we reach the conclusion {$V_{\rm{L}}(0)<0$ and $V_{\rm{R}}(0)>0$}.
When $z\rightarrow \pm\infty$, $\phi'(\pm\infty)=0$. By the analysis to Eq. (\ref{EqMz2}), we see that $A(\pm\infty )\propto -\text{ln}(\pm z)$. So $A'(\pm\infty)\propto \mp 1/z|_{z\rightarrow\infty}\rightarrow 0$ and $e^{2A(\pm\infty)}={1/z^2}|_{z\rightarrow\infty}\rightarrow 0$. Therefore, $V_{{\rm{L}}}(z)$ and $V_{{\rm{R}}}(z)$ fall off to zero at infinity.

\begin{figure*}
\includegraphics[width=0.4\textwidth]{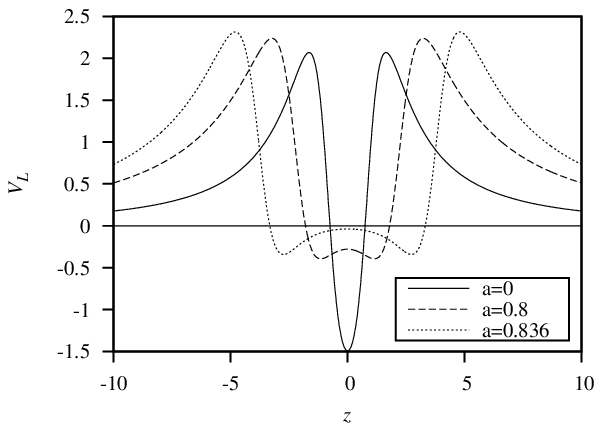}
\includegraphics[width=0.4\textwidth]{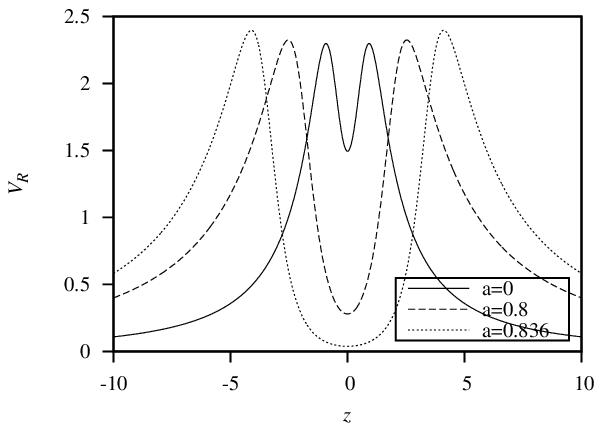}
\caption{\label{fig:potentialF} The profiles of the potentials for left- and right-handed fermions at the $z$ coordinate. The three lines correspond to $a=0, 0.8, 0.836$, respectively, and with parameters $b=2$ and $c=1$.}
\end{figure*}

\begin{figure}
\includegraphics[width=0.4\textwidth]{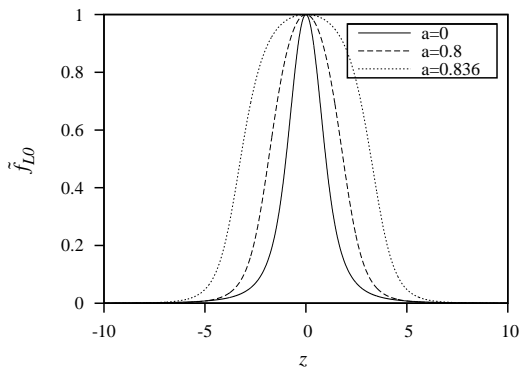}
\caption{ \label{fig:PsiZero} The profiles of the zero mode for the left-handed fermion at the $z$ coordinate. The three lines correspond to $a=0, 0.8, 0.836$, respectively. Other parameters are set to $b=2$ and $c=1$.}
\end{figure}

Equations (\ref{eq:eqmz}) have been solved numerically with the boundary conditions (\ref{bdcsz}) for $A$ and $\phi$. The
 profiles of the potentials (\ref{eq:V}) are shown in Fig.~{\ref{fig:potentialF}} for different values of $a$ and fixed $b=2$, $c=1$, and $\eta=2$. From the potential of the left-handed fermion shown in Fig.~{\ref{fig:potentialF}}, we find that, with an increase of  {the value of $a$}, both the height and width of the potential increase and there will be a double well. The minima of the potential are always lower than zero, so the left-handed fermion may have a zero mode. The zero mode is plotted in Fig.~{\ref{fig:PsiZero}} for different {values of $a$}. One can see that the width of the zero mode increases with the  {value of $a$}. For the right-handed part, there is also a potential well. As the  {value of $a$} is increasing, both the height and depth of the potential $V_{\rm{R}}(z)$ are increasing. However, there is a difference from the left-handed part, i.e., the minimum of $V_{\rm{R}}$ is always positive. So the right-handed fermion will not have a bound zero mode.

Figure~{\ref{fig:potentialF}} shows that the potentials of left- and right-handed fermions are indeed the modified volcano-type ones. Hence, there will only exist continuous spectrum for the KK modes of fermions. The volcano-type potential implies the existence of resonant or metastable states of fermions which can tunnel from the brane to the bulk~{\cite{Almeida2009,Liu2009f,Liu2009}}. In order to get the solution of massive KK modes $\tilde f_{{\rm{L}},{\rm{R}}}(z)$, we need two initial conditions for Eqs. ($\ref{eq:motionF}$) at $z=0$, which can be set as
\begin{eqnarray}
\tilde f(0)=c_1, \tilde f'(0)=0\label{even}
\end{eqnarray}
for the even parity part and
\begin{eqnarray}
\tilde f(0)=0, \tilde f'(0)=c_2\label{odd}
\end{eqnarray}
for the odd parity part, where $c_1$ and $c_2$ are arbitrary constants. Here we choose $c_1=1$ and $c_2=1$. Since Eqs. (\ref{eq:motionF}) are Schr\"{o}dinger-like we can interpret $|\tilde f_{{\rm{L}},{\rm{R}}}(z)|^2$, after normalizing $\tilde f_{{\rm{L}},{\rm{R}}}(z)$, as the probability for finding the massive KK modes on the brane. But the massive modes cannot be normalized because they are oscillating when far away from the bane along the extra dimension. Ref.~{\cite{Liu2009}} proposed a function
\begin{eqnarray}
P_{{\rm{L}},{\rm{R}}}(m^2)=
\frac{\int_{-z_b}^{z_b}|\tilde f_{{\rm{L}},{\rm{R}}}(z)|^2 dz}
     {\int_{-z_{max}}^{z_{max}}|\tilde f_{{\rm{L}},{\rm{R}}}(z)|^2 dz}
\end{eqnarray}
as the relative probability for finding the massive KK modes with mass square $m^2$ within a narrow range $-z_b<z<z_b$ around the location of a brane corresponding to a  larger interval $-z_{max}<z<z_{max}$, where $2z_b$ is  about the width of the thick brane and $z_{max}$ is set to $z_{max}=8z_b$ here. So for KK modes with $m^2\gg V_{{\rm{L}},{\rm{R}}}^{\text{max}}$, the values of $P_{{\rm{L}},{\rm{R}}}(m^2)$ will be  about $1/8$ because $f_{{\rm{L}},{\rm{R}}}$ can be approximately taken as plane waves.

\begin{figure*}
\includegraphics[width=0.4\textwidth]{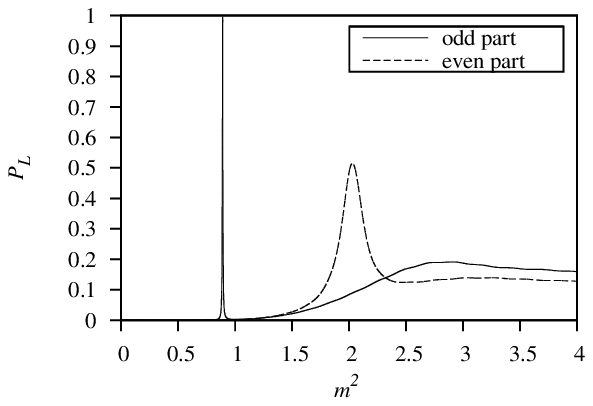}
\includegraphics[width=0.4\textwidth]{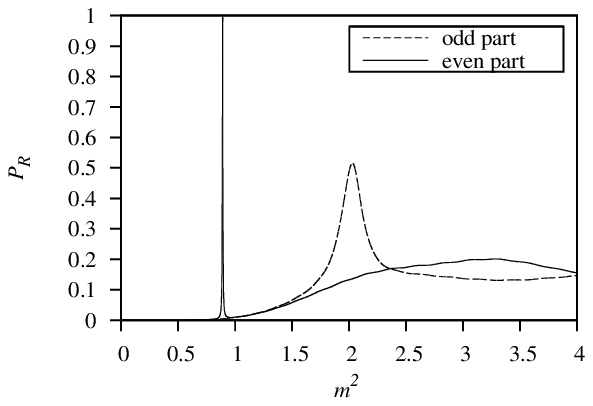}
\caption{ \label{08Spectra}The profiles of $P_{{\rm{L}},{\rm{R}}}$ with $a=0.8$, $b=2$, $c=1$, and $\eta=2$.}
\end{figure*}

\begin{figure*}
\includegraphics[width=0.4\textwidth]{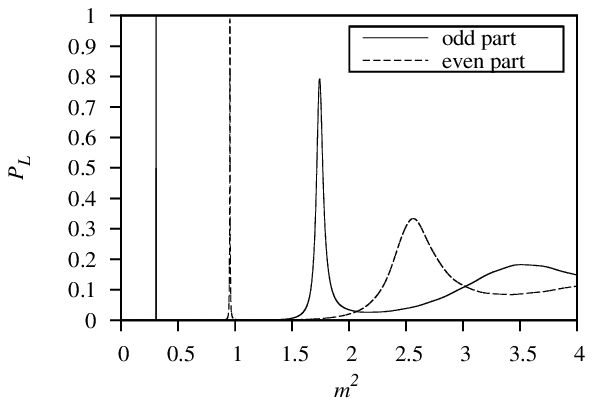}
\includegraphics[width=0.4\textwidth]{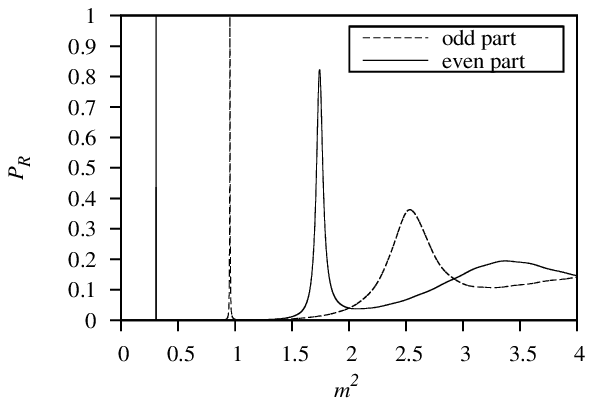}
\caption{ \label{0836Spectra}The profiles of $P_{{\rm{L}},{\rm{R}}}$ with $a=0.836$, $b=2$, $c=1$, and $\eta=2$.}
\end{figure*}

As examples, we have plotted the profiles of $P_{{\rm{L}},{\rm{R}}}(m^2)$ in  Figs.~{\ref{08Spectra}} and {\ref{0836Spectra}} corresponding to $a=0.8$ and $a=0.836$, respectively, with $b=2$ and $c=1$. From the two figures we can see that the number of resonances is the same for both the left- and right-handed fermions and increases with the  {value of $a$}. In these figures each peak corresponds to a resonant state. We can estimate the lifetime $\tau$ of every resonance as $\tau\sim \Gamma^{-1}$, where $\Gamma=\delta m$ is the width of half of the height of the corresponding peak \cite{Gregory2000}. For the case of $a=0$ there is no resonance found.

The mass $m$, width $\Gamma$, and lifetime $\tau$ of the resonances with different  {values of $a$} are listed in Table~\ref{tab3}. From the Table we can see that the masses of left- and right-handed fermion resonances are almost the same, so the formation of the massive Dirac fermions can be realized \cite{Liu2009, Almeida2009}. Table~\ref{tab3}, Fig.~{\ref{08Spectra}}, and Fig.~{\ref{0836Spectra}} also show that the increase of the  {value of $a$} will increase the number of resonances.

\begin{table*}
\begin{tabular}{|c|c|c|c|c|c|c|c|}
\hline
$a$      & Parity and Height of $V_{{\rm{L}},{\rm{R}}}$ & $n$  & $m^2$ & $m$ &$\Gamma$&$\tau$\\
\hline
       &Left-handed        &$1$& $0.307717$ & $0.554722$ & $0.00001623$&61614.2945 \\
\cline{3-7}
&$V_{\rm{L}}^{\text{max}}=2.315306$&$2$& $0.95514 $ & $0.97731$  & $0.00213339$&$468.7375$\\
\cline{3-7}
$0.836$&                &$3$& $1.74224 $ & $1.31994$  & $0.0260116\;$&38.4444\\
\cline{2-7}
          &Right-handed &$1$& $0.307716$ & $0.554722$ & $0.00001623$&$61614.2945$\\
\cline{3-7}
&$V_{\rm{R}}^{\text{max}}=2.396691$ &$2$& $0.95513 $ & $0.97731$ & $0.00209757$&$476.7421$\\
\cline{3-7}
         &             &$3$& $1.73979 $ & $1.31901$  & $0.0260411$&$38.4008$\\
\hline
       & Left-handed   &$1$& $0.890670$ & $0.943753$ & $0.002798$&$357.3981$\\
\cline{3-7}
&$V_{\rm{L}}^{\text{max}}=2.23869$    &$2$& $2.02992 $ & $1.42475$  & $0.24784$&$4.0349$\\
\cline{2-7}
$0.8$ & Right-handed  &$1$& $0.890670$ & $0.943753$ & $0.002850$&$350.877$\\
\cline{3-7}
&$V_{\rm{R}}^{\text{max}}=2.324305$  &$2$& $2.02980$ & $1.42471$ & $0.2928$&$3.4153$\\
\hline
\end{tabular}
\caption{ \label{tab3}
The mass, width, and lifetime of resonances of the left- and right-handed fermions. For the case of $a=0$ no resonance exists.  The parameters are $b=2$, $c=1$, and $\eta=2$. $V_{{\rm{L}},{\rm{R}}}^{\text{max}}$ is the maximum of the height of the potential $V_{{\rm{L}},{\rm{R}}}$, and $n$ is the order of resonant states corresponding $m^2$ from small to large. }
\end{table*}

\section{Conclusion}\label{sec:conclusion}

In this article, we have investigated the effects of  {the variation of the mass parameter of $a$} on the thick branes. We use a real scalar field, which has a potential of the $\phi^6$ model, as the background field of the thick branes.

In this model the critical  {value of $a$} of the phase transition on the brane with gravity will be lower than the one on the brane without gravity,  {and} this result is consistent with the conclusion in Ref.~\cite{Campos200288}. The discrete bound modes of fermions in the case without gravity turn into quasilocalized resonant states in the case of with gravity \cite{Davies2007}.

We find when the critical  {value of $a$} is reached the solution of the background scalar field is not unique, and it has a double-kink form. This happens in both cases with and without gravity. That means at the critical  {value of $a$} the width of the thick brane is not fixed, and with a disordered phase between two ordered phases \cite{Campos200288}. Our other main result is that the number of the bound states (in the case without gravity) or the resonant states (in the case with gravity) increases with the  {value of $a$}. This means that the branes with  {a big value of $a$} would trap fermions more efficiently.

\section*{Acknowledgments}
We would like to thank D. P. George for his sample of the {\small{GNUPLOT}} script.
This work was supported by the Program for New Century Excellent
Talents in University, the National Natural Science Foundation of
China (No. 10705013), the Doctoral Program Foundation of
Institutions of Higher Education of China (No. 20070730055 and No. 20090211110028),
the Key Project of the Chinese Ministry of Education (No. 109153),
the Natural Science Foundation of Gansu Province, China (No. 096RJZA055), and
the Fundamental Research Funds for the Central Universities (No. lzujbky-2009-54 and No. lzujbky-2009-122).

\bibliographystyle{apsrev4-1}
\bibliography{DR10549}

\begin{thebibliography}{10}%
\makeatletter
\providecommand \@ifxundefined [1]{%
 \ifx #1\undefined \expandafter \@firstoftwo
 \else \expandafter \@secondoftwo
\fi
}%
\providecommand \@ifnum [1]{%
 \ifnum #1\expandafter \@firstoftwo
 \else \expandafter \@secondoftwo
\fi
}%
\providecommand \enquote [1]{``#1''}%
\providecommand \bibnamefont  [1]{#1}%
\providecommand \bibfnamefont [1]{#1}%
\providecommand \citenamefont [1]{#1}%
\providecommand\href[0]{\@sanitize\@href}%
\providecommand\@href[1]{\endgroup\@@startlink{#1}\endgroup\@@href}%
\providecommand\@@href[1]{#1\@@endlink}%
\providecommand \@sanitize [0]{\begingroup\catcode`\&12\catcode`\#12\relax}%
\@ifxundefined \pdfoutput {\@firstoftwo}{%
 \@ifnum{\z@=\pdfoutput}{\@firstoftwo}{\@secondoftwo}%
}{%
 \providecommand\@@startlink[1]{\leavevmode\special{html:<a href="#1">}}%
 \providecommand\@@endlink[0]{\special{html:</a>}}%
}{%
 \providecommand\@@startlink[1]{%
  \leavevmode
  \pdfstartlink
   attr{/Border[0 0 1 ]/H/I/C[0 1 1]}%
   user{/Subtype/Link/A<</Type/Action/S/URI/URI(#1)>>}%
  \relax
 }%
 \providecommand\@@endlink[0]{\pdfendlink}%
}%
\providecommand \url  [0]{\begingroup\@sanitize \@url }%
\providecommand \@url [1]{\endgroup\@href {#1}{\urlprefix}}%
\providecommand \urlprefix [0]{URL }%
\providecommand \Eprint[0]{\href }%
\@ifxundefined \urlstyle {%
  \providecommand \doi [1]{doi:\discretionary{}{}{}#1}%
}{%
  \providecommand \doi [0]{doi:\discretionary{}{}{}\begingroup
  \urlstyle{rm}\Url }%
}%
\providecommand \doibase [0]{http://dx.doi.org/}%
\providecommand \Doi[1]{\href{\doibase#1}}%
\providecommand \bibAnnote [3]{%
  \BibitemShut{#1}%
  \begin{quotation}\noindent
    \textsc{Key:}\ #2\\\textsc{Annotation:}\ #3%
  \end{quotation}%
}%
\providecommand \bibAnnoteFile [2]{%
  \IfFileExists{#2}{\bibAnnote {#1} {#2} {\input{#2}}}{}%
}%
\providecommand \typeout [0]{\immediate \write \m@ne }%
\providecommand \selectlanguage [0]{\@gobble}%
\providecommand \bibinfo [0]{\@secondoftwo}%
\providecommand \bibfield [0]{\@secondoftwo}%
\providecommand \translation [1]{[#1]}%
\providecommand \BibitemOpen[0]{}%
\providecommand \bibitemStop [0]{}%
\providecommand \bibitemNoStop [0]{.\EOS\space}%
\providecommand \EOS [0]{\spacefactor3000\relax}%
\providecommand \BibitemShut [1]{\csname bibitem#1\endcsname}%
\bibitem{Antoniadis1998436}%
  \BibitemOpen
  \bibfield{author}{%
  \bibinfo {author} {\bibfnamefont{I.}~\bibnamefont{Antoniadis}}, \bibinfo
  {author} {\bibfnamefont{N.}~\bibnamefont{Arkani-Hamed}}, \bibinfo {author}
  {\bibfnamefont{S.}~\bibnamefont{Dimopoulos}},\ and\ \bibinfo {author}
  {\bibfnamefont{G.~R.}\ \bibnamefont{Dvali}},\ }%
  \bibfield{journal}{%
  \Doi{10.1016/S0370-2693(98)00860-0}{\bibinfo {journal} {Phys. Lett. B}}\ }%
  \textbf{\bibinfo {volume} {436}},\ \bibinfo {pages} {257} (\bibinfo {year}
  {1998}),\ \Eprint{http://arxiv.org/abs/hep-ph/9804398}{arXiv:hep-ph/9804398}%
  \bibAnnoteFile{NoStop}{Antoniadis1998436}%
\bibitem{Arkani-Hamed1998429}%
  \BibitemOpen
  \bibfield{author}{%
  \bibinfo {author} {\bibfnamefont{N.}~\bibnamefont{Arkani-Hamed}}, \bibinfo
  {author} {\bibfnamefont{S.}~\bibnamefont{Dimopoulos}},\ and\ \bibinfo
  {author} {\bibfnamefont{G.~R.}\ \bibnamefont{Dvali}},\ }%
  \bibfield{journal}{%
  \Doi{10.1016/S0370-2693(98)00466-3}{\bibinfo {journal} {Phys. Lett. B}}\ }%
  \textbf{\bibinfo {volume} {429}},\ \bibinfo {pages} {263} (\bibinfo {year}
  {1998}),\ \Eprint{http://arxiv.org/abs/hep-ph/9803315}{arXiv:hep-ph/9803315}%
  \bibAnnoteFile{NoStop}{Arkani-Hamed1998429}%
\bibitem{Randall199983}%
  \BibitemOpen
  \bibfield{author}{%
  \bibinfo {author} {\bibfnamefont{L.}~\bibnamefont{Randall}}\ and\ \bibinfo
  {author} {\bibfnamefont{R.}~\bibnamefont{Sundrum}},\ }%
  \bibfield{journal}{%
  \Doi{10.1103/PhysRevLett.83.3370}{\bibinfo {journal} {Phys. Rev. Lett.}}\ }%
  \textbf{\bibinfo {volume} {83}},\ \bibinfo {pages} {3370} (\bibinfo {year}
  {1999}),\ \Eprint{http://arxiv.org/abs/hep-ph/9905221}{arXiv:hep-ph/9905221}%
  \bibAnnoteFile{NoStop}{Randall199983}%
\bibitem{Rubakov1983}%
  \BibitemOpen
  \bibfield{author}{%
  \bibinfo {author} {\bibfnamefont{V.~A.}\ \bibnamefont{Rubakov}}\ and\
  \bibinfo {author} {\bibfnamefont{M.~E.}\ \bibnamefont{Shaposhnikov}},\ }%
  \bibfield{journal}{%
  \Doi{10.1016/0370-2693(83)91253-4}{\bibinfo {journal} {Phys. Lett. B}}\ }%
  \textbf{\bibinfo {volume} {125}},\ \bibinfo {pages} {136} (\bibinfo {year}
  {1983})%
  \bibAnnoteFile{NoStop}{Rubakov1983}%
\bibitem{Randall199983a}%
  \BibitemOpen
  \bibfield{author}{%
  \bibinfo {author} {\bibfnamefont{L.}~\bibnamefont{Randall}}\ and\ \bibinfo
  {author} {\bibfnamefont{R.}~\bibnamefont{Sundrum}},\ }%
  \bibfield{journal}{%
  \Doi{10.1103/PhysRevLett.83.4690}{\bibinfo {journal} {Phys. Rev. Lett.}}\ }%
  \textbf{\bibinfo {volume} {83}},\ \bibinfo {pages} {4690} (\bibinfo {year}
  {1999}),\ \Eprint{http://arxiv.org/abs/hep-th/9906064}{arXiv:hep-th/9906064}%
  \bibAnnoteFile{NoStop}{Randall199983a}%
\bibitem{Bonjour1999}%
  \BibitemOpen
  \bibfield{author}{%
  \bibinfo {author} {\bibfnamefont{F.}~\bibnamefont{Bonjour}}, \bibinfo
  {author} {\bibfnamefont{C.}~\bibnamefont{Charmousis}},\ and\ \bibinfo
  {author} {\bibfnamefont{R.}~\bibnamefont{Gregory}},\ }%
  \bibfield{journal}{%
  \Doi{10.1088/0264-9381/16/7/318}{\bibinfo {journal} {Class. Quant. Grav.}}\
  }%
  \textbf{\bibinfo {volume} {16}},\ \bibinfo {pages} {2427} (\bibinfo {year}
  {1999}),\ \Eprint{http://arxiv.org/abs/gr-qc/9902081}{arXiv:gr-qc/9902081}%
  \bibAnnoteFile{NoStop}{Bonjour1999}%
\bibitem{Goldberger1999}%
  \BibitemOpen
  \bibfield{author}{%
  \bibinfo {author} {\bibfnamefont{W.~D.}\ \bibnamefont{Goldberger}}\ and\
  \bibinfo {author} {\bibfnamefont{M.~B.}\ \bibnamefont{Wise}},\ }%
  \bibfield{journal}{%
  \Doi{10.1103/PhysRevLett.83.4922}{\bibinfo {journal} {Phys. Rev. Lett.}}\ }%
  \textbf{\bibinfo {volume} {83}},\ \bibinfo {pages} {4922} (\bibinfo {year}
  {1999}),\ \Eprint{http://arxiv.org/abs/hep-ph/9907447}{arXiv:hep-ph/9907447}%
  \bibAnnoteFile{NoStop}{Goldberger1999}%
\bibitem{Csaki2000581}%
  \BibitemOpen
  \bibfield{author}{%
  \bibinfo {author} {\bibfnamefont{C.}~\bibnamefont{Cs\'{a}ki}}, \bibinfo
  {author} {\bibfnamefont{J.}~\bibnamefont{Erlich}}, \bibinfo {author}
  {\bibfnamefont{T.~J.}\ \bibnamefont{Hollowood}},\ and\ \bibinfo {author}
  {\bibfnamefont{Y.}~\bibnamefont{Shirman}},\ }%
  \bibfield{journal}{%
  \Doi{10.1016/S0550-3213(00)00271-6}{\bibinfo {journal} {Nucl. Phys. B}}\ }%
  \textbf{\bibinfo {volume} {581}},\ \bibinfo {pages} {309} (\bibinfo {year}
  {2000}),\ \Eprint{http://arxiv.org/abs/hep-th/0001033}{arXiv:hep-th/0001033}%
  \bibAnnoteFile{NoStop}{Csaki2000581}%
\bibitem{DeWolfe200062}%
  \BibitemOpen
  \bibfield{author}{%
  \bibinfo {author} {\bibfnamefont{O.}~\bibnamefont{DeWolfe}}, \bibinfo
  {author} {\bibfnamefont{D.~Z.}\ \bibnamefont{Freedman}}, \bibinfo {author}
  {\bibfnamefont{S.~S.}\ \bibnamefont{Gubser}},\ and\ \bibinfo {author}
  {\bibfnamefont{A.}~\bibnamefont{Karch}},\ }%
  \bibfield{journal}{%
  \Doi{10.1103/PhysRevD.62.046008}{\bibinfo {journal} {Phys. Rev. D}}\ }%
  \textbf{\bibinfo {volume} {62}},\ \bibinfo {pages} {046008} (\bibinfo {year}
  {2000}),\ \Eprint{http://arxiv.org/abs/hep-th/9909134}{arXiv:hep-th/9909134}%
  \bibAnnoteFile{NoStop}{DeWolfe200062}%
\bibitem{Gremm2000478}%
  \BibitemOpen
  \bibfield{author}{%
  \bibinfo {author} {\bibfnamefont{M.}~\bibnamefont{Gremm}},\ }%
  \bibfield{journal}{%
  \Doi{10.1016/S0370-2693(00)00303-8}{\bibinfo {journal} {Phys. Lett. B}}\ }%
  \textbf{\bibinfo {volume} {478}},\ \bibinfo {pages} {434} (\bibinfo {year}
  {2000}),\ \Eprint{http://arxiv.org/abs/hep-th/9912060}{arXiv:hep-th/9912060}%
  \bibAnnoteFile{NoStop}{Gremm2000478}%
\bibitem{Ichinose200118}%
  \BibitemOpen
  \bibfield{author}{%
  \bibinfo {author} {\bibfnamefont{S.}~\bibnamefont{Ichinose}},\ }%
  \bibfield{journal}{%
  \Doi{10.1088/0264-9381/18/23/317}{\bibinfo {journal} {Class. Quant. Grav.}}\
  }%
  \textbf{\bibinfo {volume} {18}},\ \bibinfo {pages} {5239} (\bibinfo {year}
  {2001}),\ \Eprint{http://arxiv.org/abs/hep-th/0107254}{arXiv:hep-th/0107254}%
  \bibAnnoteFile{NoStop}{Ichinose200118}%
\bibitem{Ichinose200118a}%
  \BibitemOpen
  \bibfield{author}{%
  \bibinfo {author} {\bibfnamefont{S.}~\bibnamefont{Ichinose}},\ }%
  \bibfield{journal}{%
  \Doi{10.1088/0264-9381/18/3/305}{\bibinfo {journal} {Class. Quant. Grav.}}\
  }%
  \textbf{\bibinfo {volume} {18}},\ \bibinfo {pages} {421} (\bibinfo {year}
  {2001})%
  \bibAnnoteFile{NoStop}{Ichinose200118a}%
\bibitem{Kehagias2001504}%
  \BibitemOpen
  \bibfield{author}{%
  \bibinfo {author} {\bibfnamefont{A.}~\bibnamefont{Kehagias}}\ and\ \bibinfo
  {author} {\bibfnamefont{K.}~\bibnamefont{Tamvakis}},\ }%
  \bibfield{journal}{%
  \Doi{10.1016/S0370-2693(01)00274-X}{\bibinfo {journal} {Phys. Lett. B}}\ }%
  \textbf{\bibinfo {volume} {504}},\ \bibinfo {pages} {38} (\bibinfo {year}
  {2001}),\ \Eprint{http://arxiv.org/abs/hep-th/0010112}{arXiv:hep-th/0010112}%
  \bibAnnoteFile{NoStop}{Kehagias2001504}%
\bibitem{Campos200288}%
  \BibitemOpen
  \bibfield{author}{%
  \bibinfo {author} {\bibfnamefont{A.}~\bibnamefont{Campos}},\ }%
  \bibfield{journal}{%
  \Doi{10.1103/PhysRevLett.88.141602}{\bibinfo {journal} {Phys. Rev. Lett.}}\
  }%
  \textbf{\bibinfo {volume} {88}},\ \bibinfo {pages} {141602} (\bibinfo {year}
  {2002}),\ \Eprint{http://arxiv.org/abs/hep-th/0111207}{arXiv:hep-th/0111207}%
  \bibAnnoteFile{NoStop}{Campos200288}%
\bibitem{Gregory2002}%
  \BibitemOpen
  \bibfield{author}{%
  \bibinfo {author} {\bibfnamefont{R.}~\bibnamefont{Gregory}}\ and\ \bibinfo
  {author} {\bibfnamefont{A.}~\bibnamefont{Padilla}},\ }%
  \bibfield{journal}{%
  \Doi{10.1103/PhysRevD.65.084013}{\bibinfo {journal} {Phys. Rev. D}}\ }%
  \textbf{\bibinfo {volume} {65}},\ \bibinfo {pages} {084013} (\bibinfo {year}
  {2002}),\ \Eprint{http://arxiv.org/abs/hep-th/0104262}{arXiv:hep-th/0104262}%
  \bibAnnoteFile{NoStop}{Gregory2002}%
\bibitem{Gregory2002a}%
  \BibitemOpen
  \bibfield{author}{%
  \bibinfo {author} {\bibfnamefont{R.}~\bibnamefont{Gregory}}\ and\ \bibinfo
  {author} {\bibfnamefont{A.}~\bibnamefont{Padilla}},\ }%
  \bibfield{journal}{%
  \Doi{10.1088/0264-9381/19/2/308}{\bibinfo {journal} {Class. Quant. Grav.}}\
  }%
  \textbf{\bibinfo {volume} {19}},\ \bibinfo {pages} {279} (\bibinfo {year}
  {2002}),\ \Eprint{http://arxiv.org/abs/hep-th/0107108}{arXiv:hep-th/0107108}%
  \bibAnnoteFile{NoStop}{Gregory2002a}%
\bibitem{Kobayashi2002}%
  \BibitemOpen
  \bibfield{author}{%
  \bibinfo {author} {\bibfnamefont{S.}~\bibnamefont{Kobayashi}}, \bibinfo
  {author} {\bibfnamefont{K.}~\bibnamefont{Koyama}},\ and\ \bibinfo {author}
  {\bibfnamefont{J.}~\bibnamefont{Soda}},\ }%
  \bibfield{journal}{%
  \Doi{10.1103/PhysRevD.65.064014}{\bibinfo {journal} {Phys. Rev. D}}\ }%
  \textbf{\bibinfo {volume} {65}},\ \bibinfo {pages} {064014} (\bibinfo {year}
  {2002}),\ \Eprint{http://arxiv.org/abs/hep-th/0107025}{arXiv:hep-th/0107025}%
  \bibAnnoteFile{NoStop}{Kobayashi2002}%
\bibitem{Bazeia200391}%
  \BibitemOpen
  \bibfield{author}{%
  \bibinfo {author} {\bibfnamefont{D.}~\bibnamefont{Bazeia}}, \bibinfo {author}
  {\bibfnamefont{J.}~\bibnamefont{Menezes}},\ and\ \bibinfo {author}
  {\bibfnamefont{R.}~\bibnamefont{Menezes}},\ }%
  \bibfield{journal}{%
  \Doi{10.1103/PhysRevLett.91.241601}{\bibinfo {journal} {Phys. Rev. Lett.}}\
  }%
  \textbf{\bibinfo {volume} {91}},\ \bibinfo {pages} {241601} (\bibinfo {year}
  {2003})%
  \bibAnnoteFile{NoStop}{Bazeia200391}%
\bibitem{Bronnikov2003}%
  \BibitemOpen
  \bibfield{author}{%
  \bibinfo {author} {\bibfnamefont{K.~A.}\ \bibnamefont{Bronnikov}}\ and\
  \bibinfo {author} {\bibfnamefont{B.~E.}\ \bibnamefont{Meierovich}},\ }%
  \bibfield{journal}{%
  \bibinfo {journal} {Grav. Cosmol.}\ }%
  \textbf{\bibinfo {volume} {9}},\ \bibinfo {pages} {313} (\bibinfo {year}
  {2003}),\ \Eprint{http://arxiv.org/abs/gr-qc/0402030}{arXiv:gr-qc/0402030}%
  \bibAnnoteFile{NoStop}{Bronnikov2003}%
\bibitem{Eto200368}%
  \BibitemOpen
  \bibfield{author}{%
  \bibinfo {author} {\bibfnamefont{M.}~\bibnamefont{Eto}}\ and\ \bibinfo
  {author} {\bibfnamefont{N.}~\bibnamefont{Sakai}},\ }%
  \bibfield{journal}{%
  \Doi{10.1103/PhysRevD.68.125001}{\bibinfo {journal} {Phys. Rev. D}}\ }%
  \textbf{\bibinfo {volume} {68}},\ \bibinfo {pages} {125001} (\bibinfo {year}
  {2003})%
  \bibAnnoteFile{NoStop}{Eto200368}%
\bibitem{Melfo2003}%
  \BibitemOpen
  \bibfield{author}{%
  \bibinfo {author} {\bibfnamefont{A.}~\bibnamefont{Melfo}}, \bibinfo {author}
  {\bibfnamefont{N.}~\bibnamefont{Pantoja}},\ and\ \bibinfo {author}
  {\bibfnamefont{A.}~\bibnamefont{Skirzewski}},\ }%
  \bibfield{journal}{%
  \Doi{10.1103/PhysRevD.67.105003}{\bibinfo {journal} {Phys. Rev. D}}\ }%
  \textbf{\bibinfo {volume} {67}},\ \bibinfo {pages} {105003} (\bibinfo {year}
  {2003}),\ \Eprint{http://arxiv.org/abs/gr-qc/0211081}{arXiv:gr-qc/0211081}%
  \bibAnnoteFile{NoStop}{Melfo2003}%
\bibitem{Bazeia2004}%
  \BibitemOpen
  \bibfield{author}{%
  \bibinfo {author} {\bibfnamefont{D.}~\bibnamefont{Bazeia}}, \bibinfo {author}
  {\bibfnamefont{C.}~\bibnamefont{Furtado}},\ and\ \bibinfo {author}
  {\bibfnamefont{A.~R.}\ \bibnamefont{Gomes}},\ }%
  \bibfield{journal}{%
  \Doi{10.1088/1475-7516/2004/02/002}{\bibinfo {journal} {J. Cosmol. Astropart.
  Phys.}}\ }%
  \textbf{\bibinfo {volume} {0402}},\ \bibinfo {pages} {002} (\bibinfo {year}
  {2004}),\ \Eprint{http://arxiv.org/abs/hep-th/0308034}{arXiv:hep-th/0308034}%
  \bibAnnoteFile{NoStop}{Bazeia2004}%
\bibitem{Bazeia200405}%
  \BibitemOpen
  \bibfield{author}{%
  \bibinfo {author} {\bibfnamefont{D.}~\bibnamefont{Bazeia}}\ and\ \bibinfo
  {author} {\bibfnamefont{A.~R.}\ \bibnamefont{Gomes}},\ }%
  \bibfield{journal}{%
  \Doi{10.1088/1126-6708/2004/05/012}{\bibinfo {journal} {J. High Energy
  Phys.}}\ }%
  \textbf{\bibinfo {volume} {05}},\ \bibinfo {pages} {012} (\bibinfo {year}
  {2004}),\ \Eprint{http://arxiv.org/abs/hep-th/0403141}{arXiv:hep-th/0403141}%
  \bibAnnoteFile{NoStop}{Bazeia200405}%
\bibitem{Bazeia2006f}%
  \BibitemOpen
  \bibfield{author}{%
  \bibinfo {author} {\bibfnamefont{D.}~\bibnamefont{Bazeia}}, \bibinfo {author}
  {\bibfnamefont{F.~A.}\ \bibnamefont{Brito}},\ and\ \bibinfo {author}
  {\bibfnamefont{L.}~\bibnamefont{Losano}},\ }%
  \bibfield{journal}{%
  \Doi{10.1088/1126-6708/2006/11/064}{\bibinfo {journal} {J. High Energy
  Phys.}}\ }%
  \textbf{\bibinfo {volume} {11}},\ \bibinfo {pages} {064} (\bibinfo {year}
  {2006}),\ \Eprint{http://arxiv.org/abs/hep-th/0610233}{arXiv:hep-th/0610233}%
  \bibAnnoteFile{NoStop}{Bazeia2006f}%
\bibitem{Dzhunushaliev2006}%
  \BibitemOpen
  \bibfield{author}{%
  \bibinfo {author} {\bibfnamefont{V.}~\bibnamefont{Dzhunushaliev}}, \bibinfo
  {author} {\bibfnamefont{H.-J.}\ \bibnamefont{Schmidt}}, \bibinfo {author}
  {\bibfnamefont{K.}~\bibnamefont{Myrzakulov}},\ and\ \bibinfo {author}
  {\bibfnamefont{R.}~\bibnamefont{Myrzakulov}},\ }%
  \enquote{\bibinfo {title} {Thick brane solution with two scalar fields},}\
  (\bibinfo {year} {2006}),\
  \Eprint{http://arxiv.org/abs/gr-qc/0610100}{arXiv:gr-qc/0610100}%
  \bibAnnoteFile{NoStop}{Dzhunushaliev2006}%
\bibitem{Dzhunushaliev200713}%
  \BibitemOpen
  \bibfield{author}{%
  \bibinfo {author} {\bibfnamefont{V.}~\bibnamefont{Dzhunushaliev}},\ }%
  \bibfield{journal}{%
  \bibinfo {journal} {Grav. Cosmol.}\ }%
  \textbf{\bibinfo {volume} {13}},\ \bibinfo {pages} {302} (\bibinfo {year}
  {2007}),\ \Eprint{http://arxiv.org/abs/gr-qc/0603020}{arXiv:gr-qc/0603020}%
  \bibAnnoteFile{NoStop}{Dzhunushaliev200713}%
\bibitem{George2007}%
  \BibitemOpen
  \bibfield{author}{%
  \bibinfo {author} {\bibfnamefont{D.~P.}\ \bibnamefont{George}}\ and\ \bibinfo
  {author} {\bibfnamefont{R.~R.}\ \bibnamefont{Volkas}},\ }%
  \bibfield{journal}{%
  \Doi{10.1103/PhysRevD.75.105007}{\bibinfo {journal} {Phys. Rev. D}}\ }%
  \textbf{\bibinfo {volume} {75}},\ \bibinfo {pages} {105007} (\bibinfo {year}
  {2007})%
  \bibAnnoteFile{NoStop}{George2007}%
\bibitem{Dzhunushaliev200877}%
  \BibitemOpen
  \bibfield{author}{%
  \bibinfo {author} {\bibfnamefont{V.}~\bibnamefont{Dzhunushaliev}}, \bibinfo
  {author} {\bibfnamefont{V.}~\bibnamefont{Folomeev}}, \bibinfo {author}
  {\bibfnamefont{D.}~\bibnamefont{Singleton}},\ and\ \bibinfo {author}
  {\bibfnamefont{S.}~\bibnamefont{Aguilar-Rudametkin}},\ }%
  \bibfield{journal}{%
  \Doi{10.1103/PhysRevD.77.044006}{\bibinfo {journal} {Phys. Rev. D}}\ }%
  \textbf{\bibinfo {volume} {77}},\ \bibinfo {pages} {044006} (\bibinfo {year}
  {2008}),\ \Eprint{http://arxiv.org/abs/hep-th/0703043}{arXiv:hep-th/0703043}%
  \bibAnnoteFile{NoStop}{Dzhunushaliev200877}%
\bibitem{Bazeia2009c}%
  \BibitemOpen
  \bibfield{author}{%
  \bibinfo {author} {\bibfnamefont{D.}~\bibnamefont{Bazeia}}, \bibinfo {author}
  {\bibfnamefont{A.~R.}\ \bibnamefont{Gomes}}, \bibinfo {author}
  {\bibfnamefont{L.}~\bibnamefont{Losano}},\ and\ \bibinfo {author}
  {\bibfnamefont{R.}~\bibnamefont{Menezes}},\ }%
  \bibfield{journal}{%
  \Doi{10.1016/j.physletb.2008.12.039}{\bibinfo {journal} {Phys. Lett. B}}\ }%
  \textbf{\bibinfo {volume} {671}},\ \bibinfo {pages} {402} (\bibinfo {year}
  {2009}),\ \Eprint{http://arxiv.org/abs/0808.1815}{arXiv:0808.1815 [hep-th]}%
  \bibAnnoteFile{NoStop}{Bazeia2009c}%
\bibitem{Burnier2009}%
  \BibitemOpen
  \bibfield{author}{%
  \bibinfo {author} {\bibfnamefont{Y.}~\bibnamefont{Burnier}}\ and\ \bibinfo
  {author} {\bibfnamefont{K.}~\bibnamefont{Zuleta}},\ }%
  \bibfield{journal}{%
  \bibinfo {journal} {J. High Energy Phys.}\ }%
  \textbf{\bibinfo {volume} {2009}},\ \bibinfo {pages} {065} (\bibinfo {year}
  {2009})%
  \bibAnnoteFile{NoStop}{Burnier2009}%
\bibitem{Chumbes2010}%
  \BibitemOpen
  \bibfield{author}{%
  \bibinfo {author} {\bibfnamefont{A.~E.~R.}\ \bibnamefont{Chumbes}}\ and\
  \bibinfo {author} {\bibfnamefont{M.~B.}\ \bibnamefont{Hott}},\ }%
  \bibfield{journal}{%
  \Doi{10.1103/PhysRevD.81.045008}{\bibinfo {journal} {Phys. Rev. D}}\ }%
  \textbf{\bibinfo {volume} {81}},\ \bibinfo {pages} {045008} (\bibinfo {year}
  {2010}),\ \Eprint{http://arxiv.org/abs/0905.4715}{arXiv:0905.4715 [hep-th]}%
  \bibAnnoteFile{NoStop}{Chumbes2010}%
\bibitem{Dzhunushaliev200979}%
  \BibitemOpen
  \bibfield{author}{%
  \bibinfo {author} {\bibfnamefont{V.}~\bibnamefont{Dzhunushaliev}}, \bibinfo
  {author} {\bibfnamefont{V.}~\bibnamefont{Folomeev}},\ and\ \bibinfo {author}
  {\bibfnamefont{M.}~\bibnamefont{Minamitsuji}},\ }%
  \bibfield{journal}{%
  \Doi{10.1103/PhysRevD.79.024001}{\bibinfo {journal} {Phys. Rev. D}}\ }%
  \textbf{\bibinfo {volume} {79}},\ \bibinfo {pages} {024001} (\bibinfo {year}
  {2009}),\ \Eprint{http://arxiv.org/abs/0809.4076}{arXiv:0809.4076 [gr-qc]}%
  \bibAnnoteFile{NoStop}{Dzhunushaliev200979}%
\bibitem{Liu2009c}%
  \BibitemOpen
  \bibfield{author}{%
  \bibinfo {author} {\bibfnamefont{Y.-X.}\ \bibnamefont{Liu}}, \bibinfo
  {author} {\bibfnamefont{Y.}~\bibnamefont{Zhong}},\ and\ \bibinfo {author}
  {\bibfnamefont{K.}~\bibnamefont{Yang}},\ }%
  \bibfield{journal}{%
  \Doi{10.1209/0295-5075/90/51001}{\bibinfo {journal} {Europhys. Lett.}}\ }%
  \textbf{\bibinfo {volume} {90}},\ \bibinfo {pages} {51001} (\bibinfo {year}
  {2010}),\ \Eprint{http://arxiv.org/abs/0907.1952}{arXiv:0907.1952 [hep-th]}%
  \bibAnnoteFile{NoStop}{Liu2009c}%
\bibitem{Arias2002643}%
  \BibitemOpen
  \bibfield{author}{%
  \bibinfo {author} {\bibfnamefont{O.}~\bibnamefont{Arias}}, \bibinfo {author}
  {\bibfnamefont{R.}~\bibnamefont{Cardenas}},\ and\ \bibinfo {author}
  {\bibfnamefont{I.}~\bibnamefont{Quiros}},\ }%
  \bibfield{journal}{%
  \Doi{10.1016/S0550-3213(02)00691-0}{\bibinfo {journal} {Nucl. Phys. B}}\ }%
  \textbf{\bibinfo {volume} {643}},\ \bibinfo {pages} {187} (\bibinfo {year}
  {2002}),\ \Eprint{http://arxiv.org/abs/hep-th/0202130}{arXiv:hep-th/0202130}%
  \bibAnnoteFile{NoStop}{Arias2002643}%
\bibitem{Barbosa-Cendejas200510}%
  \BibitemOpen
  \bibfield{author}{%
  \bibinfo {author} {\bibfnamefont{N.}~\bibnamefont{Barbosa-Cendejas}}\ and\
  \bibinfo {author} {\bibfnamefont{A.}~\bibnamefont{Herrera-Aguilar}},\ }%
  \bibfield{journal}{%
  \bibinfo {journal} {J. High Energy Phys.}\ }%
  \textbf{\bibinfo {volume} {10}},\ \bibinfo {pages} {101} (\bibinfo {year}
  {2005}),\ \Eprint{http://arxiv.org/abs/hep-th/0511050}{arXiv:hep-th/0511050}%
  \bibAnnoteFile{NoStop}{Barbosa-Cendejas200510}%
\bibitem{Barbosa-Cendejas200673}%
  \BibitemOpen
  \bibfield{author}{%
  \bibinfo {author} {\bibfnamefont{N.}~\bibnamefont{Barbosa-Cendejas}}\ and\
  \bibinfo {author} {\bibfnamefont{A.}~\bibnamefont{Herrera-Aguilar}},\ }%
  \bibfield{journal}{%
  \Doi{10.1103/PhysRevD.73.084022}{\bibinfo {journal} {Phys. Rev. D}}\ }%
  \textbf{\bibinfo {volume} {73}},\ \bibinfo {pages} {084022} (\bibinfo {year}
  {2006}),\ \Eprint{http://arxiv.org/abs/hep-th/0603184}{arXiv:hep-th/0603184}%
  \bibAnnoteFile{NoStop}{Barbosa-Cendejas200673}%
\bibitem{Barbosa-Cendejas2007}%
  \BibitemOpen
  \bibfield{author}{%
  \bibinfo {author} {\bibfnamefont{N.}~\bibnamefont{Barbosa-Cendejas}},
  \bibinfo {author} {\bibfnamefont{A.}~\bibnamefont{Herrera-Aguilar}}, \bibinfo
  {author} {\bibfnamefont{U.}~\bibnamefont{Nucamendi}},\ and\ \bibinfo {author}
  {\bibfnamefont{I.}~\bibnamefont{Quiros}},\ }%
  \enquote{\bibinfo {title} {Mass hierarchy and mass gap on thick branes with
  poincare symmetry},}\  (\bibinfo {year} {2007}),\
  \Eprint{http://arxiv.org/abs/0712.3098}{arXiv:0712.3098 [hep-th]}%
  \bibAnnoteFile{NoStop}{Barbosa-Cendejas2007}%
\bibitem{Barbosa-Cendejas200877}%
  \BibitemOpen
  \bibfield{author}{%
  \bibinfo {author} {\bibfnamefont{N.}~\bibnamefont{Barbosa-Cendejas}},
  \bibinfo {author} {\bibfnamefont{A.}~\bibnamefont{Herrera-Aguilar}}, \bibinfo
  {author} {\bibfnamefont{M.~A.}\ \bibnamefont{Reyes~Santos}},\ and\ \bibinfo
  {author} {\bibfnamefont{C.}~\bibnamefont{Schubert}},\ }%
  \bibfield{journal}{%
  \Doi{10.1103/PhysRevD.77.126013}{\bibinfo {journal} {Phys. Rev. D}}\ }%
  \textbf{\bibinfo {volume} {77}},\ \bibinfo {pages} {126013} (\bibinfo {year}
  {2008}),\ \Eprint{http://arxiv.org/abs/0709.3552}{arXiv:0709.3552 [hep-th]}%
  \bibAnnoteFile{NoStop}{Barbosa-Cendejas200877}%
\bibitem{Liu2009g}%
  \BibitemOpen
  \bibfield{author}{%
  \bibinfo {author} {\bibfnamefont{Y.-X.}\ \bibnamefont{Liu}}, \bibinfo
  {author} {\bibfnamefont{K.}~\bibnamefont{Yang}},\ and\ \bibinfo {author}
  {\bibfnamefont{Y.}~\bibnamefont{Zhong}},\ }%
  \enquote{\bibinfo {title} {de sitter thick brane solution in weyl
  geometry},}\  (\bibinfo {year} {2009}),\
  \Eprint{http://arxiv.org/abs/0911.0269}{arXiv:0911.0269 [hep-th]}%
  \bibAnnoteFile{NoStop}{Liu2009g}%
\bibitem{Andrianov2003}%
  \BibitemOpen
  \bibfield{author}{%
  \bibinfo {author} {\bibfnamefont{A.~A.}\ \bibnamefont{Andrianov}}, \bibinfo
  {author} {\bibfnamefont{V.~A.}\ \bibnamefont{Andrianov}}, \bibinfo {author}
  {\bibfnamefont{P.}~\bibnamefont{Giacconi}},\ and\ \bibinfo {author}
  {\bibfnamefont{R.}~\bibnamefont{Soldati}},\ }%
  \bibfield{journal}{%
  \bibinfo {journal} {J. High Energy Phys.}\ }%
  \textbf{\bibinfo {volume} {07}},\ \bibinfo {pages} {063} (\bibinfo {year}
  {2003}),\ \Eprint{http://arxiv.org/abs/hep-ph/0305271}{arXiv:hep-ph/0305271}%
  \bibAnnoteFile{NoStop}{Andrianov2003}%
\bibitem{Andrianov2005}%
  \BibitemOpen
  \bibfield{author}{%
  \bibinfo {author} {\bibfnamefont{A.~A.}\ \bibnamefont{Andrianov}}, \bibinfo
  {author} {\bibfnamefont{V.~A.}\ \bibnamefont{Andrianov}}, \bibinfo {author}
  {\bibfnamefont{P.}~\bibnamefont{Giacconi}},\ and\ \bibinfo {author}
  {\bibfnamefont{R.}~\bibnamefont{Soldati}},\ }%
  \bibfield{journal}{%
  \bibinfo {journal} {J. High Energy Phys.}\ }%
  \textbf{\bibinfo {volume} {07}},\ \bibinfo {pages} {003} (\bibinfo {year}
  {2005}),\ \Eprint{http://arxiv.org/abs/hep-th/0503115}{arXiv:hep-th/0503115}%
  \bibAnnoteFile{NoStop}{Andrianov2005}%
\bibitem{SenGupta2008}%
  \BibitemOpen
  \bibfield{author}{%
  \bibinfo {author} {\bibfnamefont{S.}~\bibnamefont{SenGupta}},\ }%
  \enquote{\bibinfo {title} {Aspects of warped braneworld models},}\  (\bibinfo
  {year} {2008}),\ \Eprint{http://arxiv.org/abs/0812.1092}{arXiv:0812.1092
  [hep-th]}%
  \bibAnnoteFile{NoStop}{SenGupta2008}%
\bibitem{Dzhunushaliev2009}%
  \BibitemOpen
  \bibfield{author}{%
  \bibinfo {author} {\bibfnamefont{V.}~\bibnamefont{Dzhunushaliev}}, \bibinfo
  {author} {\bibfnamefont{V.}~\bibnamefont{Folomeev}},\ and\ \bibinfo {author}
  {\bibfnamefont{M.}~\bibnamefont{Minamitsuji}},\ }%
  \bibfield{journal}{%
  \Doi{10.1088/0034-4885/73/6/066901}{\bibinfo {journal} {Rept. Prog. Phys.}}\
  }%
  \textbf{\bibinfo {volume} {73}},\ \bibinfo {pages} {066901} (\bibinfo {year}
  {2010}),\ \Eprint{http://arxiv.org/abs/0904.1775}{arXiv:0904.1775 [gr-qc]}%
  \bibAnnoteFile{NoStop}{Dzhunushaliev2009}%
\bibitem{Shifman2009}%
  \BibitemOpen
  \bibfield{author}{%
  \bibinfo {author} {\bibfnamefont{M.}~\bibnamefont{Shifman}},\ }%
  \bibfield{journal}{%
  \Doi{10.1142/S0217751X10048548}{\bibinfo {journal} {Rept. Prog. Phys.}}\ }%
  \textbf{\bibinfo {volume} {73}},\ \bibinfo {pages} {066901} (\bibinfo {year}
  {2009}),\ \Eprint{http://arxiv.org/abs/0907.3074}{arXiv:0907.3074 [hep-ph]}%
  \bibAnnoteFile{NoStop}{Shifman2009}%
\bibitem{Rizzo2010}%
  \BibitemOpen
  \bibfield{author}{%
  \bibinfo {author} {\bibfnamefont{T.~G.}\ \bibnamefont{Rizzo}},\ }%
  \bibfield{journal}{%
  \Doi{10.1063/1.3473866}{\bibinfo {journal} {AIP Conf. Proc.}}\ }%
  \textbf{\bibinfo {volume} {1256}},\ \bibinfo {pages} {27} (\bibinfo {year}
  {2010}),\ \Eprint{http://arxiv.org/abs/1003.1698}{arXiv:1003.1698 [hep-ph]}%
  \bibAnnoteFile{NoStop}{Rizzo2010}%
\bibitem{Bajc2000}%
  \BibitemOpen
  \bibfield{author}{%
  \bibinfo {author} {\bibfnamefont{B.}~\bibnamefont{Bajc}}\ and\ \bibinfo
  {author} {\bibfnamefont{G.}~\bibnamefont{Gabadadze}},\ }%
  \bibfield{journal}{%
  \Doi{10.1016/S0370-2693(00)00055-1}{\bibinfo {journal} {Phys. Lett. B}}\ }%
  \textbf{\bibinfo {volume} {474}},\ \bibinfo {pages} {282} (\bibinfo {year}
  {2000}),\ \Eprint{http://arxiv.org/abs/hep-th/9912232}{arXiv:hep-th/9912232}%
  \bibAnnoteFile{NoStop}{Bajc2000}%
\bibitem{Randjbar-Daemi2000}%
  \BibitemOpen
  \bibfield{author}{%
  \bibinfo {author} {\bibfnamefont{S.}~\bibnamefont{Randjbar-Daemi}}\ and\
  \bibinfo {author} {\bibfnamefont{M.~E.}\ \bibnamefont{Shaposhnikov}},\ }%
  \bibfield{journal}{%
  \Doi{10.1016/S0370-2693(00)01100-X}{\bibinfo {journal} {Phys. Lett. B}}\ }%
  \textbf{\bibinfo {volume} {492}},\ \bibinfo {pages} {361} (\bibinfo {year}
  {2000}),\ \Eprint{http://arxiv.org/abs/hep-th/0008079}{arXiv:hep-th/0008079}%
  \bibAnnoteFile{NoStop}{Randjbar-Daemi2000}%
\bibitem{Pomarol2000}%
  \BibitemOpen
  \bibfield{author}{%
  \bibinfo {author} {\bibfnamefont{A.}~\bibnamefont{Pomarol}},\ }%
  \bibfield{journal}{%
  \Doi{10.1016/S0370-2693(00)00737-1}{\bibinfo {journal} {Phys. Lett. B}}\ }%
  \textbf{\bibinfo {volume} {486}},\ \bibinfo {pages} {153} (\bibinfo {year}
  {2000}),\ \Eprint{http://arxiv.org/abs/hep-ph/9911294}{arXiv:hep-ph/9911294}%
  \bibAnnoteFile{NoStop}{Pomarol2000}%
\bibitem{Oda2001}%
  \BibitemOpen
  \bibfield{author}{%
  \bibinfo {author} {\bibfnamefont{I.}~\bibnamefont{Oda}},\ }%
  \enquote{\bibinfo {title} {A new mechanism for trapping of photon},}\
  (\bibinfo {year} {2001}),\
  \Eprint{http://arxiv.org/abs/hep-th/0103052}{arXiv:hep-th/0103052}%
  \bibAnnoteFile{NoStop}{Oda2001}%
\bibitem{Akhmedov2001}%
  \BibitemOpen
  \bibfield{author}{%
  \bibinfo {author} {\bibfnamefont{E.~K.}\ \bibnamefont{Akhmedov}},\ }%
  \bibfield{journal}{%
  \Doi{10.1016/S0370-2693(01)01176-5}{\bibinfo {journal} {Phys. Lett. B}}\ }%
  \textbf{\bibinfo {volume} {521}},\ \bibinfo {pages} {79} (\bibinfo {year}
  {2001}),\ \Eprint{http://arxiv.org/abs/hep-th/0107223}{arXiv:hep-th/0107223}%
  \bibAnnoteFile{NoStop}{Akhmedov2001}%
\bibitem{Dvali2001}%
  \BibitemOpen
  \bibfield{author}{%
  \bibinfo {author} {\bibfnamefont{G.~R.}\ \bibnamefont{Dvali}}, \bibinfo
  {author} {\bibfnamefont{G.}~\bibnamefont{Gabadadze}},\ and\ \bibinfo {author}
  {\bibfnamefont{M.~A.}\ \bibnamefont{Shifman}},\ }%
  \bibfield{journal}{%
  \Doi{10.1016/S0370-2693(00)01329-0}{\bibinfo {journal} {Phys. Lett. B}}\ }%
  \textbf{\bibinfo {volume} {497}},\ \bibinfo {pages} {271} (\bibinfo {year}
  {2001}),\ \Eprint{http://arxiv.org/abs/hep-th/0010071}{arXiv:hep-th/0010071}%
  \bibAnnoteFile{NoStop}{Dvali2001}%
\bibitem{Ghoroku2002}%
  \BibitemOpen
  \bibfield{author}{%
  \bibinfo {author} {\bibfnamefont{K.}~\bibnamefont{Ghoroku}}\ and\ \bibinfo
  {author} {\bibfnamefont{A.}~\bibnamefont{Nakamura}},\ }%
  \bibfield{journal}{%
  \Doi{10.1103/PhysRevD.65.084017}{\bibinfo {journal} {Phys. Rev. D}}\ }%
  \textbf{\bibinfo {volume} {65}},\ \bibinfo {pages} {084017} (\bibinfo {year}
  {2002}),\ \Eprint{http://arxiv.org/abs/hep-th/0106145}{arXiv:hep-th/0106145}%
  \bibAnnoteFile{NoStop}{Ghoroku2002}%
\bibitem{Ringeval200265}%
  \BibitemOpen
  \bibfield{author}{%
  \bibinfo {author} {\bibfnamefont{C.}~\bibnamefont{Ringeval}}, \bibinfo
  {author} {\bibfnamefont{P.}~\bibnamefont{Peter}},\ and\ \bibinfo {author}
  {\bibfnamefont{J.-P.}\ \bibnamefont{Uzan}},\ }%
  \bibfield{journal}{%
  \Doi{10.1103/PhysRevD.65.044016}{\bibinfo {journal} {Phys. Rev. D}}\ }%
  \textbf{\bibinfo {volume} {65}},\ \bibinfo {pages} {044016} (\bibinfo {year}
  {2002}),\ \Eprint{http://arxiv.org/abs/hep-th/0109194}{arXiv:hep-th/0109194}%
  \bibAnnoteFile{NoStop}{Ringeval200265}%
\bibitem{Koley200522}%
  \BibitemOpen
  \bibfield{author}{%
  \bibinfo {author} {\bibfnamefont{R.}~\bibnamefont{Koley}}\ and\ \bibinfo
  {author} {\bibfnamefont{S.}~\bibnamefont{Kar}},\ }%
  \bibfield{journal}{%
  \Doi{10.1088/0264-9381/22/4/008}{\bibinfo {journal} {Class. Quant. Grav.}}\
  }%
  \textbf{\bibinfo {volume} {22}},\ \bibinfo {pages} {753} (\bibinfo {year}
  {2005}),\ \Eprint{http://arxiv.org/abs/hep-th/0407158}{arXiv:hep-th/0407158}%
  \bibAnnoteFile{NoStop}{Koley200522}%
\bibitem{Wang2005}%
  \BibitemOpen
  \bibfield{author}{%
  \bibinfo {author} {\bibfnamefont{Y.-Q.}\ \bibnamefont{Wang}}, \bibinfo
  {author} {\bibfnamefont{T.-Y.}\ \bibnamefont{Si}}, \bibinfo {author}
  {\bibfnamefont{Y.-X.}\ \bibnamefont{Liu}},\ and\ \bibinfo {author}
  {\bibfnamefont{Y.-S.}\ \bibnamefont{Duan}},\ }%
  \bibfield{journal}{%
  \Doi{10.1142/S0217732305018037}{\bibinfo {journal} {Mod. Phys. Lett. A}}\ }%
  \textbf{\bibinfo {volume} {20}},\ \bibinfo {pages} {3045} (\bibinfo {year}
  {2005}),\ \Eprint{http://arxiv.org/abs/hep-th/0508111}{arXiv:hep-th/0508111}%
  \bibAnnoteFile{NoStop}{Wang2005}%
\bibitem{Melfo2006}%
  \BibitemOpen
  \bibfield{author}{%
  \bibinfo {author} {\bibfnamefont{A.}~\bibnamefont{Melfo}}, \bibinfo {author}
  {\bibfnamefont{N.}~\bibnamefont{Pantoja}},\ and\ \bibinfo {author}
  {\bibfnamefont{J.~D.}\ \bibnamefont{Tempo}},\ }%
  \bibfield{journal}{%
  \Doi{10.1103/PhysRevD.73.044033}{\bibinfo {journal} {Phys. Rev. D}}\ }%
  \textbf{\bibinfo {volume} {73}},\ \bibinfo {pages} {044033} (\bibinfo {year}
  {2006}),\ \Eprint{http://arxiv.org/abs/hep-th/0601161}{arXiv:hep-th/0601161}%
  \bibAnnoteFile{NoStop}{Melfo2006}%
\bibitem{Liu2007a}%
  \BibitemOpen
  \bibfield{author}{%
  \bibinfo {author} {\bibfnamefont{Y.-X.}\ \bibnamefont{Liu}}, \bibinfo
  {author} {\bibfnamefont{L.}~\bibnamefont{Zhao}}, \bibinfo {author}
  {\bibfnamefont{X.-H.}\ \bibnamefont{Zhang}},\ and\ \bibinfo {author}
  {\bibfnamefont{Y.-S.}\ \bibnamefont{Duan}},\ }%
  \bibfield{journal}{%
  \Doi{10.1016/j.nuclphysb.2007.05.018}{\bibinfo {journal} {Nucl. Phys. B}}\ }%
  \textbf{\bibinfo {volume} {785}},\ \bibinfo {pages} {234} (\bibinfo {year}
  {2007}),\ \Eprint{http://arxiv.org/abs/0704.2812}{arXiv:0704.2812 [hep-th]}%
  \bibAnnoteFile{NoStop}{Liu2007a}%
\bibitem{Liu2007}%
  \BibitemOpen
  \bibfield{author}{%
  \bibinfo {author} {\bibfnamefont{Y.-X.}\ \bibnamefont{Liu}}, \bibinfo
  {author} {\bibfnamefont{L.}~\bibnamefont{Zhao}},\ and\ \bibinfo {author}
  {\bibfnamefont{Y.-S.}\ \bibnamefont{Duan}},\ }%
  \bibfield{journal}{%
  \Doi{10.1088/1126-6708/2007/04/097}{\bibinfo {journal} {J. High Energy
  Phys.}}\ }%
  \textbf{\bibinfo {volume} {04}},\ \bibinfo {pages} {097} (\bibinfo {year}
  {2007}),\ \Eprint{http://arxiv.org/abs/hep-th/0701010}{arXiv:hep-th/0701010}%
  \bibAnnoteFile{NoStop}{Liu2007}%
\bibitem{Davies2007}%
  \BibitemOpen
  \bibfield{author}{%
  \bibinfo {author} {\bibfnamefont{R.}~\bibnamefont{Davies}}\ and\ \bibinfo
  {author} {\bibfnamefont{D.~P.}\ \bibnamefont{George}},\ }%
  \bibfield{journal}{%
  \Doi{10.1103/PhysRevD.76.104010}{\bibinfo {journal} {Phys. Rev. D}}\ }%
  \textbf{\bibinfo {volume} {76}},\ \bibinfo {pages} {104010} (\bibinfo {year}
  {2007}),\ \Eprint{http://arxiv.org/abs/0705.1391}{arXiv:0705.1391 [hep-ph]}%
  \bibAnnoteFile{NoStop}{Davies2007}%
\bibitem{Slatyer2007}%
  \BibitemOpen
  \bibfield{author}{%
  \bibinfo {author} {\bibfnamefont{T.~R.}\ \bibnamefont{Slatyer}}\ and\
  \bibinfo {author} {\bibfnamefont{R.~R.}\ \bibnamefont{Volkas}},\ }%
  \bibfield{journal}{%
  \Doi{10.1088/1126-6708/2007/04/062}{\bibinfo {journal} {J. High Energy
  Phys.}}\ }%
  \textbf{\bibinfo {volume} {04}},\ \bibinfo {pages} {062} (\bibinfo {year}
  {2007}),\ \Eprint{http://arxiv.org/abs/hep-ph/0609003}{arXiv:hep-ph/0609003}%
  \bibAnnoteFile{NoStop}{Slatyer2007}%
\bibitem{Koley2008}%
  \BibitemOpen
  \bibfield{author}{%
  \bibinfo {author} {\bibfnamefont{R.}~\bibnamefont{Koley}},\ }%
  \enquote{\bibinfo {title} {Localization of fields on brane},}\  (\bibinfo
  {year} {2008}),\ \Eprint{http://arxiv.org/abs/0812.1423}{arXiv:0812.1423
  [hep-th]}%
  \bibAnnoteFile{NoStop}{Koley2008}%
\bibitem{Zhang2008}%
  \BibitemOpen
  \bibfield{author}{%
  \bibinfo {author} {\bibfnamefont{X.-H.}\ \bibnamefont{Zhang}}, \bibinfo
  {author} {\bibfnamefont{Y.-X.}\ \bibnamefont{Liu}},\ and\ \bibinfo {author}
  {\bibfnamefont{Y.-S.}\ \bibnamefont{Duan}},\ }%
  \bibfield{journal}{%
  \Doi{10.1142/S0217732308026133}{\bibinfo {journal} {Mod. Phys. Lett. A}}\ }%
  \textbf{\bibinfo {volume} {23}},\ \bibinfo {pages} {2093} (\bibinfo {year}
  {2008}),\ \Eprint{http://arxiv.org/abs/0709.1888}{arXiv:0709.1888 [hep-th]}%
  \bibAnnoteFile{NoStop}{Zhang2008}%
\bibitem{Liu200878}%
  \BibitemOpen
  \bibfield{author}{%
  \bibinfo {author} {\bibfnamefont{Y.-X.}\ \bibnamefont{Liu}}, \bibinfo
  {author} {\bibfnamefont{L.-D.}\ \bibnamefont{Zhang}}, \bibinfo {author}
  {\bibfnamefont{L.-J.}\ \bibnamefont{Zhang}},\ and\ \bibinfo {author}
  {\bibfnamefont{Y.-S.}\ \bibnamefont{Duan}},\ }%
  \bibfield{journal}{%
  \Doi{10.1103/PhysRevD.78.065025}{\bibinfo {journal} {Phys. Rev. D}}\ }%
  \textbf{\bibinfo {volume} {78}},\ \bibinfo {pages} {065025} (\bibinfo {year}
  {2008}),\ \Eprint{http://arxiv.org/abs/0804.4553}{arXiv:0804.4553 [hep-th]}%
  \bibAnnoteFile{NoStop}{Liu200878}%
\bibitem{Liu200808}%
  \BibitemOpen
  \bibfield{author}{%
  \bibinfo {author} {\bibfnamefont{Y.-X.}\ \bibnamefont{Liu}}, \bibinfo
  {author} {\bibfnamefont{L.-D.}\ \bibnamefont{Zhang}}, \bibinfo {author}
  {\bibfnamefont{S.-W.}\ \bibnamefont{Wei}},\ and\ \bibinfo {author}
  {\bibfnamefont{Y.-S.}\ \bibnamefont{Duan}},\ }%
  \bibfield{journal}{%
  \Doi{10.1088/1126-6708/2008/08/041}{\bibinfo {journal} {J. High Energy
  Phys.}}\ }%
  \textbf{\bibinfo {volume} {08}},\ \bibinfo {pages} {041} (\bibinfo {year}
  {2008}),\ \Eprint{http://arxiv.org/abs/0803.0098}{arXiv:0803.0098 [hep-th]}%
  \bibAnnoteFile{NoStop}{Liu200808}%
\bibitem{Liu200802}%
  \BibitemOpen
  \bibfield{author}{%
  \bibinfo {author} {\bibfnamefont{Y.-X.}\ \bibnamefont{Liu}}, \bibinfo
  {author} {\bibfnamefont{X.-H.}\ \bibnamefont{Zhang}}, \bibinfo {author}
  {\bibfnamefont{L.-D.}\ \bibnamefont{Zhang}},\ and\ \bibinfo {author}
  {\bibfnamefont{Y.-S.}\ \bibnamefont{Duan}},\ }%
  \bibfield{journal}{%
  \Doi{10.1088/1126-6708/2008/02/067}{\bibinfo {journal} {J. High Energy
  Phys.}}\ }%
  \textbf{\bibinfo {volume} {02}},\ \bibinfo {pages} {067} (\bibinfo {year}
  {2008}),\ \Eprint{http://arxiv.org/abs/0708.0065}{arXiv:0708.0065 [hep-th]}%
  \bibAnnoteFile{NoStop}{Liu200802}%
\bibitem{Guerrero2009}%
  \BibitemOpen
  \bibfield{author}{%
  \bibinfo {author} {\bibfnamefont{R.}~\bibnamefont{Guerrero}}, \bibinfo
  {author} {\bibfnamefont{A.}~\bibnamefont{Melfo}}, \bibinfo {author}
  {\bibfnamefont{N.}~\bibnamefont{Pantoja}},\ and\ \bibinfo {author}
  {\bibfnamefont{R.~O.}\ \bibnamefont{Rodriguez}},\ }%
  \bibfield{journal}{%
  \Doi{10.1103/PhysRevD.81.086004}{\bibinfo {journal} {Phys. Rev. D}}\ }%
  \textbf{\bibinfo {volume} {81}},\ \bibinfo {pages} {086004} (\bibinfo {year}
  {2010}),\ \Eprint{http://arxiv.org/abs/0912.0463}{arXiv:0912.0463 [hep-th]}%
  \bibAnnoteFile{NoStop}{Guerrero2009}%
\bibitem{Zhao2009}%
  \BibitemOpen
  \bibfield{author}{%
  \bibinfo {author} {\bibfnamefont{Z.-H.}\ \bibnamefont{Zhao}}, \bibinfo
  {author} {\bibfnamefont{Y.-X.}\ \bibnamefont{Liu}},\ and\ \bibinfo {author}
  {\bibfnamefont{H.-T.}\ \bibnamefont{Li}},\ }%
  \bibfield{journal}{%
  \Doi{10.1088/0264-9381/27/18/185001}{\bibinfo {journal} {Class. Quant.
  Grav.}}\ }%
  \textbf{\bibinfo {volume} {27}},\ \bibinfo {pages} {185001} (\bibinfo {year}
  {2010}),\ \Eprint{http://arxiv.org/abs/0911.2572}{arXiv:0911.2572 [hep-th]}%
  \bibAnnoteFile{NoStop}{Zhao2009}%
\bibitem{Liu2009a}%
  \BibitemOpen
  \bibfield{author}{%
  \bibinfo {author} {\bibfnamefont{Y.-X.}\ \bibnamefont{Liu}}, \bibinfo
  {author} {\bibfnamefont{C.-E.}\ \bibnamefont{Fu}}, \bibinfo {author}
  {\bibfnamefont{L.}~\bibnamefont{Zhao}},\ and\ \bibinfo {author}
  {\bibfnamefont{Y.-S.}\ \bibnamefont{Duan}},\ }%
  \bibfield{journal}{%
  \bibinfo {journal} {Phys. Rev. D}\ }%
  \textbf{\bibinfo {volume} {80}},\ \bibinfo {pages} {065020} (\bibinfo {year}
  {2009}),\ \Eprint{http://arxiv.org/abs/0907.0910}{arXiv:0907.0910 [hep-th]}%
  \bibAnnoteFile{NoStop}{Liu2009a}%
\bibitem{Liu2010}%
  \BibitemOpen
  \bibfield{author}{%
  \bibinfo {author} {\bibfnamefont{Y.-X.}\ \bibnamefont{Liu}}, \bibinfo
  {author} {\bibfnamefont{H.}~\bibnamefont{Guo}}, \bibinfo {author}
  {\bibfnamefont{C.-E.}\ \bibnamefont{Fu}},\ and\ \bibinfo {author}
  {\bibfnamefont{J.-R.}\ \bibnamefont{Ren}},\ }%
  \bibfield{journal}{%
  \bibinfo {journal} {J. High Energy Phys.}\ }%
  \textbf{\bibinfo {volume} {02}},\ \bibinfo {pages} {080} (\bibinfo {year}
  {2010}),\ \Eprint{http://arxiv.org/abs/0907.4424}{arXiv:0907.4424 [hep-th]}%
  \bibAnnoteFile{NoStop}{Liu2010}%
\bibitem{Dubovsky200062}%
  \BibitemOpen
  \bibfield{author}{%
  \bibinfo {author} {\bibfnamefont{S.~L.}\ \bibnamefont{Dubovsky}}, \bibinfo
  {author} {\bibfnamefont{V.~A.}\ \bibnamefont{Rubakov}},\ and\ \bibinfo
  {author} {\bibfnamefont{P.~G.}\ \bibnamefont{Tinyakov}},\ }%
  \bibfield{journal}{%
  \Doi{10.1103/PhysRevD.62.105011}{\bibinfo {journal} {Phys. Rev. D}}\ }%
  \textbf{\bibinfo {volume} {62}},\ \bibinfo {pages} {105011} (\bibinfo {year}
  {2000}),\ \Eprint{http://arxiv.org/abs/hep-th/0006046}{arXiv:hep-th/0006046}%
  \bibAnnoteFile{NoStop}{Dubovsky200062}%
\bibitem{Almeida2009}%
  \BibitemOpen
  \bibfield{author}{%
  \bibinfo {author} {\bibfnamefont{C.~A.~S.}\ \bibnamefont{Almeida}}, \bibinfo
  {author} {\bibfnamefont{R.}~\bibnamefont{Casana}}, \bibinfo {author}
  {\bibnamefont{\text{M. M. Ferreira, Jr.}}},\ and\ \bibinfo {author}
  {\bibfnamefont{A.~R.}\ \bibnamefont{Gomes}},\ }%
  \bibfield{journal}{%
  \Doi{10.1103/PhysRevD.79.125022}{\bibinfo {journal} {Phys. Rev. D}}\ }%
  \textbf{\bibinfo {volume} {79}},\ \bibinfo {pages} {125022} (\bibinfo {year}
  {2009}),\ \Eprint{http://arxiv.org/abs/0901.3543}{arXiv:0901.3543 [hep-th]}%
  \bibAnnoteFile{NoStop}{Almeida2009}%
\bibitem{Liu2009}%
  \BibitemOpen
  \bibfield{author}{%
  \bibinfo {author} {\bibfnamefont{Y.-X.}\ \bibnamefont{Liu}}, \bibinfo
  {author} {\bibfnamefont{J.}~\bibnamefont{Yang}}, \bibinfo {author}
  {\bibfnamefont{Z.-H.}\ \bibnamefont{Zhao}}, \bibinfo {author}
  {\bibfnamefont{C.-E.}\ \bibnamefont{Fu}},\ and\ \bibinfo {author}
  {\bibfnamefont{Y.-S.}\ \bibnamefont{Duan}},\ }%
  \bibfield{journal}{%
  \Doi{10.1103/PhysRevD.80.065019}{\bibinfo {journal} {Phys. Rev. D}}\ }%
  \textbf{\bibinfo {volume} {80}},\ \bibinfo {pages} {065019} (\bibinfo {year}
  {2009}),\ \Eprint{http://arxiv.org/abs/0904.1785}{arXiv:0904.1785 [hep-th]}%
  \bibAnnoteFile{NoStop}{Liu2009}%
\bibitem{Liu2009f}%
  \BibitemOpen
  \bibfield{author}{%
  \bibinfo {author} {\bibfnamefont{Y.-X.}\ \bibnamefont{Liu}}, \bibinfo
  {author} {\bibfnamefont{H.-T.}\ \bibnamefont{Li}}, \bibinfo {author}
  {\bibfnamefont{Z.-H.}\ \bibnamefont{Zhao}}, \bibinfo {author}
  {\bibfnamefont{J.-X.}\ \bibnamefont{Li}},\ and\ \bibinfo {author}
  {\bibfnamefont{J.-R.}\ \bibnamefont{Ren}},\ }%
  \bibfield{journal}{%
  \Doi{10.1088/1126-6708/2009/10/091}{\bibinfo {journal} {J. High Energy
  Phys.}}\ }%
  \textbf{\bibinfo {volume} {10}},\ \bibinfo {pages} {091} (\bibinfo {year}
  {2009}),\ \Eprint{http://arxiv.org/abs/0909.2312}{arXiv:0909.2312 [hep-th]}%
  \bibAnnoteFile{NoStop}{Liu2009f}%
\bibitem{Liu2010a}%
  \BibitemOpen
  \bibfield{author}{%
  \bibinfo {author} {\bibfnamefont{Y.-X.}\ \bibnamefont{Liu}}, \bibinfo
  {author} {\bibfnamefont{C.-E.}\ \bibnamefont{Fu}}, \bibinfo {author}
  {\bibfnamefont{H.}~\bibnamefont{Guo}}, \bibinfo {author}
  {\bibfnamefont{S.-W.}\ \bibnamefont{Wei}},\ and\ \bibinfo {author}
  {\bibfnamefont{Z.-H.}\ \bibnamefont{Zhao}},\ }%
  \enquote{\bibinfo {title} {Bulk matters on a grs-inspired braneworld},}\
  (\bibinfo {year} {2010}),\
  \Eprint{http://arxiv.org/abs/1002.2130}{arXiv:1002.2130 [hep-th]}%
  \bibAnnoteFile{NoStop}{Liu2010a}%
\bibitem{Dolan1974}%
  \BibitemOpen
  \bibfield{author}{%
  \bibinfo {author} {\bibfnamefont{L.}~\bibnamefont{Dolan}}\ and\ \bibinfo
  {author} {\bibfnamefont{R.}~\bibnamefont{Jackiw}},\ }%
  \bibfield{journal}{%
  \Doi{10.1103/PhysRevD.9.3320}{\bibinfo {journal} {Phys. Rev. D}}\ }%
  \textbf{\bibinfo {volume} {9}},\ \bibinfo {pages} {3320} (\bibinfo {year}
  {1974})%
  \bibAnnoteFile{NoStop}{Dolan1974}%
\bibitem{Weinberg1974}%
  \BibitemOpen
  \bibfield{author}{%
  \bibinfo {author} {\bibfnamefont{S.}~\bibnamefont{Weinberg}},\ }%
  \bibfield{journal}{%
  \Doi{10.1103/PhysRevD.9.3357}{\bibinfo {journal} {Phys. Rev. D}}\ }%
  \textbf{\bibinfo {volume} {9}},\ \bibinfo {pages} {3357} (\bibinfo {month}
  {Jun}\ \bibinfo {year} {1974})%
  \bibAnnoteFile{NoStop}{Weinberg1974}%
\bibitem{Jackiw1975}%
  \BibitemOpen
  \bibfield{author}{%
  \bibinfo {author} {\bibfnamefont{R.}~\bibnamefont{Jackiw}},\ }%
  \bibfield{journal}{%
  \Doi{10.1007/BFb0013349}{\bibinfo {journal} {Lecture Notes in Physics}}\ }%
  \textbf{\bibinfo {volume} {39}},\ \bibinfo {pages} {319} (\bibinfo {year}
  {1975})%
  \bibAnnoteFile{NoStop}{Jackiw1975}%
\bibitem{Joseph1982}%
  \BibitemOpen
  \bibfield{author}{%
  \bibinfo {author} {\bibfnamefont{K.~B.}\ \bibnamefont{Joseph}}\ and\ \bibinfo
  {author} {\bibfnamefont{V.~C.}\ \bibnamefont{Kuriakose}},\ }%
  \bibfield{journal}{%
  \bibinfo {journal} {J. Phys. A: Math. Gen.}\ }%
  \textbf{\bibinfo {volume} {15}},\ \bibinfo {pages} {2231} (\bibinfo {year}
  {1982})%
  \bibAnnoteFile{NoStop}{Joseph1982}%
\bibitem{Su198720}%
  \BibitemOpen
  \bibfield{author}{%
  \bibinfo {author} {\bibfnamefont{R.}~\bibnamefont{Su}}\ and\ \bibinfo
  {author} {\bibfnamefont{T.}~\bibnamefont{Chen}},\ }%
  \bibfield{journal}{%
  \bibinfo {journal} {J. Phys. A: Math. Gen.}\ }%
  \textbf{\bibinfo {volume} {20}},\ \bibinfo {pages} {5939} (\bibinfo {year}
  {1987})%
  \bibAnnoteFile{NoStop}{Su198720}%
\bibitem{Brevik2001599}%
  \BibitemOpen
  \bibfield{author}{%
  \bibinfo {author} {\bibfnamefont{I.~H.}\ \bibnamefont{Brevik}}, \bibinfo
  {author} {\bibfnamefont{K.~A.}\ \bibnamefont{Milton}}, \bibinfo {author}
  {\bibfnamefont{S.}~\bibnamefont{Nojiri}},\ and\ \bibinfo {author}
  {\bibfnamefont{S.~D.}\ \bibnamefont{Odintsov}},\ }%
  \bibfield{journal}{%
  \Doi{10.1016/S0550-3213(01)00026-8}{\bibinfo {journal} {Nucl. Phys. B}}\ }%
  \textbf{\bibinfo {volume} {599}},\ \bibinfo {pages} {305} (\bibinfo {year}
  {2001}),\ \Eprint{http://arxiv.org/abs/hep-th/0010205}{arXiv:hep-th/0010205}%
  \bibAnnoteFile{NoStop}{Brevik2001599}%
\bibitem{Bazeia200411}%
  \BibitemOpen
  \bibfield{author}{%
  \bibinfo {author} {\bibfnamefont{D.}~\bibnamefont{Bazeia}}, \bibinfo {author}
  {\bibfnamefont{F.~A.}\ \bibnamefont{Brito}},\ and\ \bibinfo {author}
  {\bibfnamefont{A.~R.}\ \bibnamefont{Gomes}},\ }%
  \bibfield{journal}{%
  \Doi{10.1088/1126-6708/2004/11/070}{\bibinfo {journal} {J. High Energy
  Phys.}}\ }%
  \textbf{\bibinfo {volume} {11}},\ \bibinfo {pages} {070} (\bibinfo {year}
  {2004}),\ \Eprint{http://arxiv.org/abs/hep-th/0411088}{arXiv:hep-th/0411088}%
  \bibAnnoteFile{NoStop}{Bazeia200411}%
\bibitem{Note1}%
  \BibitemOpen
  \bibinfo {note} {Because Eq. (\ref {Eq:Phi}) is non-linear, so in order to
  solve it numerically we need to offer an initial guess solution of $\phi $.
  And different guess solutions follow different solutions of $\phi $.}%
  \bibAnnoteFile{Stop}{Note1}%
\bibitem{Duan1958}%
  \BibitemOpen
  \bibfield{author}{%
  \bibinfo {author} {\bibfnamefont{I.}~\bibnamefont{Duan}},\ }%
  \bibfield{journal}{%
  \bibinfo {journal} {J. Exp. Theor. Phys.}\ }%
  \textbf{\bibinfo {volume} {34}},\ \bibinfo {pages} {632} (\bibinfo {year}
  {1958})%
  \bibAnnoteFile{NoStop}{Duan1958}%
\bibitem{Fischbach1981}%
  \BibitemOpen
  \bibfield{author}{%
  \bibinfo {author} {\bibfnamefont{E.}~\bibnamefont{Fischbach}}, \bibinfo
  {author} {\bibfnamefont{B.~S.}\ \bibnamefont{Freeman}},\ and\ \bibinfo
  {author} {\bibfnamefont{W.-K.}\ \bibnamefont{Cheng}},\ }%
  \bibfield{journal}{%
  \Doi{10.1103/PhysRevD.23.2157}{\bibinfo {journal} {Phys. Rev. D}}\ }%
  \textbf{\bibinfo {volume} {23}},\ \bibinfo {pages} {2157} (\bibinfo {year}
  {1981})%
  \bibAnnoteFile{NoStop}{Fischbach1981}%
\bibitem{Zhao200776}%
  \BibitemOpen
  \bibfield{author}{%
  \bibinfo {author} {\bibfnamefont{Z.-H.}\ \bibnamefont{Zhao}}, \bibinfo
  {author} {\bibfnamefont{Y.-X.}\ \bibnamefont{Liu}},\ and\ \bibinfo {author}
  {\bibfnamefont{X.-G.}\ \bibnamefont{Lee}},\ }%
  \bibfield{journal}{%
  \Doi{10.1103/PhysRevD.76.064016}{\bibinfo {journal} {Phys. Rev. D}}\ }%
  \textbf{\bibinfo {volume} {76}},\ \bibinfo {pages} {064016} (\bibinfo {year}
  {2007})%
  \bibAnnoteFile{NoStop}{Zhao200776}%
\bibitem{Gregory2000}%
  \BibitemOpen
  \bibfield{author}{%
  \bibinfo {author} {\bibfnamefont{R.}~\bibnamefont{Gregory}}, \bibinfo
  {author} {\bibfnamefont{V.~A.}\ \bibnamefont{Rubakov}},\ and\ \bibinfo
  {author} {\bibfnamefont{S.~M.}\ \bibnamefont{Sibiryakov}},\ }%
  \bibfield{journal}{%
  \Doi{10.1103/PhysRevLett.84.5928}{\bibinfo {journal} {Phys. Rev. Lett.}}\ }%
  \textbf{\bibinfo {volume} {84}},\ \bibinfo {pages} {5928} (\bibinfo {year}
  {2000}),\ \Eprint{http://arxiv.org/abs/hep-th/0002072}{arXiv:hep-th/0002072}%
  \bibAnnoteFile{NoStop}{Gregory2000}%
\end{thebibliography}%
\end{document}